\documentclass [11pt]{article}
\pdfoutput=1
\usepackage{amsfonts}
\usepackage{soul}
\usepackage{graphicx}
\usepackage{amsmath,amssymb, dsfont}
\usepackage{latexsym}
\usepackage{mathrsfs}
\usepackage{bbold}
\usepackage{color}
\usepackage{slashed}
\usepackage{cancel}
\usepackage{soul} %
\usepackage{setspace} %
\usepackage{comment}
\usepackage{booktabs}
\usepackage{multirow}
\usepackage{cite}
\usepackage[small]{caption}

\definecolor{red}{rgb}{1,0,0}

\makeatletter
\def\section{\@startsection {section}{1}{\z@}{-3.5ex plus -1ex minus
 -.2ex}{2.3ex plus .2ex}{\large\bf}}
\def\subsection{\@startsection{subsection}{2}{\z@}{-3.25ex plus -1ex
minus -.2ex}{1.5ex plus .2ex}{\normalsize\bf}}
\makeatother
\makeatletter

\@addtoreset{equation}{section}

\makeatother

\newcommand{\bea}{\begin{equation} \begin{aligned}} \newcommand{\eea}{\end{aligned} \end{equation}}
\def\be{\begin{equation}} \def\ee{\end{equation}} 
\def\nn{\nonumber}

\usepackage[colorlinks,linkcolor=black,citecolor=blue,urlcolor=blue,linktocpage,pagebackref]{hyperref}

\allowdisplaybreaks

\begin{document}

\thispagestyle{empty}

\begin{center}

	\vspace*{-.6cm}

	\begin{center}

		\vspace*{1.1cm}

		{\centering \Large\textbf{Looking through the QCD Conformal Window\\[5pt] with Perturbation Theory}}

	\end{center}

	\vspace{0.8cm}
	{\bf Lorenzo Di Pietro$^{a,b}$ and Marco Serone$^{b,c,d}$}

	\vspace{1.cm}
	
	${}^a\!\!$
	{\em  Dipartimento di Fisica, Universit\`a di Trieste, \\ Strada Costiera 11, I-34151 Trieste, Italy}
		
	\vspace{.3cm}

	${}^b\!\!$
	{\em INFN, Sezione di Trieste, Via Valerio 2, I-34127 Trieste, Italy}

	\vspace{.3cm}

	${}^c\!\!$
	{\em SISSA, Via Bonomea 265, I-34136 Trieste, Italy}

	\vspace{.3cm}

	${}^d\!\!$
	{\em ICTP, Strada Costiera 11, I-34151 Trieste, Italy}

	\vspace{.3cm}

\end{center}

\vspace{1cm}

\centerline{\bf Abstract}
\vspace{2 mm}
\begin{quote}

We study the conformal window of QCD using perturbation theory, starting from the perturbative upper edge and going down as much as we can towards the strongly coupled regime.
We do so by exploiting the available five-loop computation of the $\overline{{\rm MS}}$ $\beta$-function and employing Borel resummation techniques both for the ordinary perturbative series and for the Banks-Zaks conformal expansion. Large-$n_f$ results are also used. We argue that the perturbative series for the $\overline{{\rm MS}}$ $\beta$-function is most likely asymptotic and {\it non}-Borel resummable, yet Borel resummation techniques allow to improve on ordinary perturbation theory. 
We find substantial evidence that QCD with $n_f=12$ flavours flows in the IR to a conformal field theory.
Though the evidence is weaker, we find indications that also $n_f=11$ might sit within the conformal window. 
We also compute the anomalous dimensions $\gamma$ and $\gamma_g$ of respectively the fermion mass bilinear and the gauge kinetic term operator at the fixed point, and compare them with the available lattice results.
The conformal window might extend for lower values of $n_f$, but our methods break down for $n_f < 11$, where we expect that non-perturbative effects become important. 
A similar analysis is performed in the Veneziano limit.

\end{quote}

\newpage

\tableofcontents

\section{Introduction}

Understanding the IR behavior of four-dimensional non-abelian gauge theories is one of the most interesting problems in high energy physics. 
Even at zero temperature and chemical potential, this behavior can be drastically different for different choices of the gauge group and matter content.
In particular, UV-free gauge theories with matter admit so-called conformal windows, regions in parameter space where 
they flow to non-trivial conformal field theories (CFTs) in the IR. The most notable example is given by 
$SU(n_c)$ gauge theories with $n_f$ fermions in the fundamental representation of the gauge group.
At fixed $n_c$, the conformal window spans an interval \mbox{$n_f^{*} \leq n_f \leq 11 n_c/2$}. The upper edge of the conformal window, where $n_f \approx 11n_c/2$, 
is easily studied because it is accessible in perturbation theory. 
Determining the lower edge of the conformal window $n_f^*$ is instead a non-trivial task.

Computations based on (uncontrolled) approximations of Schwinger-Dyson gap equations suggest that $n_f^*\approx 11.9$ for $n_c=3$ and
$x^* \approx 4$ in the Veneziano limit where $n_c,n_f\rightarrow \infty$, with $x=n_f/n_c$ fixed \cite{Appelquist:1996dq}.
These results also indicate that at the lower edge of the conformal window the bilinear fermion operator $\bar \psi \psi$ acquires an anomalous dimension $\gamma \approx -1$ \cite{Yamawaki:1985zg,Appelquist:1988yc,Cohen:1988sq}. Computations based on (uncontrolled) truncations of exact Renormalization Group (RG) flow equations give $n_f^* \approx 10$ for $n_c=3$   \cite{Gies:2005as}.
Similar computations, making also use of bi-furcation theory, give $x^* \approx 4$ in the Veneziano limit \cite{Gukov:2016tnp,Kuipers:2018lux}
and $n_f^*\approx 12$  for $n_c=3$  \cite{Kuipers:2018lux} (see also \cite{Kusafuka:2011fd}).  A phenomenological holographic bottom-up model of non-abelian gauge theories in the Veneziano limit suggests that $3.7 \lesssim x^* \lesssim 4.2$ \cite{Jarvinen:2011qe} (see also \cite{Alvares:2012kr}). Previous analysis based on the perturbative series in the gauge-coupling  found $n_f^*\approx  9$ \cite{Ryttov:2016ner, Antipin:2018asc} for $n_c=3$, and using the Banks-Zaks conformal expansion found $n_f^* \approx 10$\cite{Kim:2020yvr} or $n_f^* \approx 9$ \cite{Ryttov:2017lkz} for $n_c=3$, and $x^* \approx 2.9$ \cite{Ryttov:2017lkz} in the Veneziano limit.\footnote{Given our estimates of the errors due to both the numerical extrapolation and the non-perturbative corrections, we think it is too optimistic to hope to reach the lower edge of the conformal window using only few coefficients in perturbation theory as an input.\label{footnoteintro}}

Lattice methods are the only ones so far based on first principles.\footnote{The conformal bootstrap (see e.g. \cite{Simmons-Duffin:2016gjk} for an introduction and \cite{Poland:2018epd} for an extensive review 
oriented on numerical results) is also a viable first-principle method alternative to the lattice. However, CFTs expected to arise in the IR from 
non-abelian gauge theories coupled to fermion matter sit well within the allowed region of CFTs, while the numerical techniques developed so far allow us to study only CFTs living at the edges of the allowed regions. We hope that further progress would allow us in the future to use the conformal bootstrap to study the conformal window of four-dimensional non-abelian gauge theories.}
Studying a non-abelian gauge theory in its conformal phase on the lattice is a hard task and there is no consensus yet on the value of $n_f^*$ for QCD ($n_c=3$).
More specifically, the lattice community did not reach a unanimous consensus on the case $n_f=12$, which is reported to be within the QCD conformal window by most groups, with the exception
of one (see \cite{DeGrand:2015zxa}, in particular table I, for a review and an extensive list of references). The controversy on the $n_f=12$ case is still open, see e.g. \cite{Hasenfratz:2018wpq,Fodor:2018uih,Hasenfratz:2019dpr} for some of the latest works that appeared after \cite{DeGrand:2015zxa}.\footnote{For completeness, most lattice studies indicate that $n_f=8$ is outside the conformal window, while the case $n_f=10$ is unclear.}

The aim of this paper is to study the conformal window of gauge theories using perturbation theory, starting from the upper edge and going down as much as we can towards the strongly coupled regime.
We will do so by exploiting the recent remarkable five-loop computation of the $\beta$-function in $\overline{{\rm MS}}$  \cite{Baikov:2016tgj,Herzog:2017ohr,Luthe:2017ttg,Chetyrkin:2017bjc} and by employing Borel resummation techniques both for the ordinary perturbative series and for a Banks-Zaks conformal expansion \cite{Banks:1981nn}, namely an expansion in $\epsilon\propto11n_c/2-n_f$ along the one-parameter family of CFTs starting from the upper edge of the conformal window. We will also make use of results in the large-$n_f$ limit of non-abelian gauge theories.  In this limit the theory of course is no longer UV-free and no IR fixed point is expected, yet
comparing the known exact results at the first non-trivial order in $1/n_f$ with the predictions coming from our resummations provide a useful sanity check and a guidance for the numerics.

We find substantial evidence that $n_f=12$ is within the conformal window of QCD. The value of $\gamma$ at the fixed point seems to indicate that the theory there is still relatively weakly coupled.
Though the evidence is weaker, we find indications that also $n_f=11$ sits within the conformal window. The conformal window might extend for lower values of $n_f$, but our methods break down for $n_f<11$, where we expect that non-perturbative effects become important. In the context of the scenario advocated in \cite{Kaplan:2009kr} other couplings are expected to approach marginality, and the study of the RG flow of the single gauge coupling might be too restrictive. 
A similar analysis is performed in the Veneziano limit, giving evidence that the conformal window extends at least up to $x=4.2$.

 In section \ref{sec:nature} we set the stage for our analysis, starting with a brief review of basic facts about perturbative expansions in non-abelian gauge theories.
The nature of the ordinary perturbative expansion of  $\beta$ and $\gamma$ in non-abelian gauge theories
is only known in the large-$n_f$ limit at fixed $n_c$. In this case both $\beta$ and $\gamma$ admit, at each known order in the $1/n_f$ expansion, a convergent series expansion in $\overline{{\rm MS}}$, in contrast to what happens in other more physical renormalization schemes, such as momentum subtraction or on-shell \cite{Broadhurst:1992si}.  Surprisingly enough (at least to us), at fixed $n_f$ and $n_c$ or in the Veneziano limit, it is not known whether the $\overline{{\rm MS}}$ perturbative series for $\beta$ and $\gamma$  is convergent or divergent asymptotic. We will argue that the most plausible case is the second one, and that most likely the series is not Borel resummable.
We then briefly review basic facts of the conformal Banks-Zaks expansion. The first terms of the conformal series indicate a better behaviour than that of the ordinary coupling series, 
though the conformal series should also be divergent asymptotic under the above assumption on the gauge coupling expansion.

In section 3 we will clarify what it means and to what extent it is useful to attempt to Borel resum a perturbative series that is expected to be non-Borel resummable.
We will see that in doing so one could and in fact does get an improvement with respect to perturbation theory, that as we will see turns out to be crucial for the $n_f=12$ case of QCD, where 
perturbation theory shows a fixed point up to 4-loops, but not at 5-loops.

In section 4 we finally present our results based on numerical resummations of the perturbative series using Pad\'e-Borel approximants. We first use the exact large-$n_f$ results to improve on some aspects of the numerics and then we report the numerically resummed $\beta$-function as a function of the coupling in the Veneziano limit for some values of $x$ and for QCD for different values of $n_f$.
We compare the values of the coupling constant at the fixed point with those obtained by performing a conformal expansion in $\epsilon$. The agreement between the two procedures provides a sanity check of the procedure. 
As further check, in addition to the anomalous dimension $\gamma$ of $\bar \psi \psi$, we also study  the anomalous dimension $\gamma_g$ of the gauge kinetic term operator ${\rm Tr}[ F_{\mu\nu}F^{\mu\nu}]$.
Our findings for $\gamma$ and $\gamma_g$ are in good agreement
with the available lattice results for $n_f=12$.

We conclude in section 5.  Some further details are relegated in four appendices. In particular, we collect the perturbative 5-loop result for $\beta$ and $\gamma$ computed in the literature in appendix  \ref{app:betagamma}, we review the exact large-$n_f$ computation of $\gamma$ in appendix  \ref{app:largenf}, we report some details of the numerical Borel resummation in appendix \ref{app:numerics}
and we review some known mathematical facts about convergence properties of Pad\'e approximants in appendix \ref{app:pade}. The last appendix can be read independently of the rest of the paper.

\paragraph{Note Added (December 2024):} The previous version of this work had a typo of a factor of 2 in the coefficient $\kappa$ in eq. \eqref{errorNP}, resulting in a significant underestimate of the non-perturbative contribution to the error. For $n_f=12$, the total error estimate is substantially affected in the gauge coupling expansion while the 
conformal expansion is not significantly modified. 
To reach the same qualitative conclusions as in the previous version for the coupling constant expansion, we change the (arbitrary) coefficient from the very conservative choice $c_{\text{np}}=10$ to the value $c_{\text{np}}=0.1$ used here. However, since the central values and the conformal expansion results are unaffected, we believe that the evidence for the existence of the fixed point with $n_f=12$ remains good and essentially unmodified. The present results indicate however that the conformal expansion is significantly more accurate than the coupling expansion. 
We thank Fabiana De Cesare for spotting this typo.

\section{Nature of Perturbative Series for RG Functions}
\label{sec:nature}

The knowledge of the QCD $\beta$-function with $n_f$ fundamental fermions has recently been extended up to five-loop orders \cite{Baikov:2016tgj} in $\overline{{\rm MS}}$.
The result of \cite{Baikov:2016tgj} has been verified and extended to more general non-abelian gauge theories with fermions in \cite{Herzog:2017ohr,Luthe:2017ttg,Chetyrkin:2017bjc}.
Similarly the five-loop fermion mass anomalous dimension for QCD was obtained in \cite{Baikov:2014qja} and extended to a generic gauge group and fermion representation in \cite{Luthe:2016xec,Baikov:2017ujl}.
Using the notation of \cite{Herzog:2017ohr}, we denote the loopwise expansion parameter 
\be
a \equiv \frac{g^2}{16\pi^2}
\ee
where $g$ is the usual gauge coupling constant. The general expressions for $\beta(a)$ and $\gamma(a)$ can be found in the above references, e.g. $\beta$ in eqs.(3.1)-(3.5) of \cite{Herzog:2017ohr},
and $\gamma$ in eqs.(4.1)-(4.3) of \cite{Baikov:2017ujl}.
For the reader's convenience we report in appendix \ref{app:betagamma} the expression of $\beta(a)$ and $\gamma(a)$ 
as a function of $n_f$ in QCD, and $\beta(\lambda)$  and $\gamma(\lambda)$ as a function of $x$ in the Veneziano limit,
$n_c\rightarrow \infty$, $n_f\rightarrow\infty$, with $x\equiv n_f/n_c$ and $\lambda \equiv a n_c$ held fixed.
The large order behaviour of the perturbative series of $\beta$ and $\gamma$ in non-abelian gauge theories is largely unknown. We briefly review what is known below and then explain our expectations for such large order behaviour.

\subsection{Gauge Coupling Expansion}\label{sec:gce}

The perturbative gauge coupling expansion of physical observables in non-abelian gauge theories is generally divergent asymptotic and non-Borel resummable because of the presence of non-perturbative effects.
These typically give rise to singularities located at real positive values of the argument of the Borel transform of observables, making the Laplace transform needed to recover the original observable ill-defined. 
Some of these singularities are associated to genuine semi-classical configurations, ordinary gauge instantons. In euclidean space, the need of integrating around instanton configurations,
real classical finite action solutions, immediately signal the impossibility of having a Borel resummable perturbative series. Other singularities, apparently not associated to semi-classical configurations, 
are related to the so called renormalons, see ref.\cite{Beneke:1998ui} for a review.  While instanton singularities are associated to the factorial proliferation of Feynman diagrams in QFT, 
renormalons are related to a specific set of Feynman diagrams that give a factorially growing contribution to the perturbative series.
More generally, it is a common lore that renormalons are expected to appear in the perturbative expansion of physical observables in a classically marginal and log-running coupling.\footnote{See however \cite{Marino:2019fvu} for an exception in two dimensions.}
 
The nature of the ordinary perturbative series for RG functions such as $\beta$ and $\gamma$ has been established only in the large-$n_f$ limit, i.e. in an expansion in $1/n_f$ with both $n_c$ and the 't~Hooft-like coupling $\lambda_f \equiv n_f a$ held fixed.
Remarkably, in this limit the ${\cal O}(1/n_f)$ terms in $\beta(\lambda_f)$ and $\gamma(\lambda_f)$ can be computed exactly. 
Since $\beta$ and $\gamma$ are not physical observables away from fixed points, their form is renormalization-scheme dependent. In $\overline{{\rm MS}}$ their expression is given in eq.s~\eqref{eq:largenfbeta} and \eqref{eq:largenfgamma}.
Notably, the ${\cal O}(1/n_f)$ terms of  $\beta$ and $\gamma$ are analytic functions of the coupling constant in a neighbourhood of $\lambda_f=0$, implying that the perturbative expansion for these RG functions has a finite radius of convergence in the large-$n_f$ limit. In particular, the contribution of the bubble diagrams that dominate in the large-$n_f$ limit does not grow factorially and hence they do not give rise to renormalons. 
The ${\cal O}(1/n_f)$ terms of the $\beta$-function have also been studied in more physical mass-dependent renormalization schemes, such as momentum subtraction or on-shell scheme, for QED \cite{Broadhurst:1992si}.
In contrast to $\overline{{\rm MS}}$, bubble diagrams have been shown to display a factorial growth in these cases, leading eventually to a divergent asymptotic expansions for the ${\cal O}(1/n_f)$ terms of $\beta$ in the large-$n_f$ limit. 
The good large-$n_f$ behaviour of the $\overline{\rm MS}$ $\beta$-function has been considered an indication that, somewhat surprisingly, it might admit a convergent perturbative expansion at finite $n_f$ (see the question I in the Conclusion of the review \cite{Beneke:1998ui}).

In the rest of this section we would like to make a few observations about the nature of the perturbative expansion for $\beta$ and $\gamma$ in $\overline{\rm MS}$  beyond the large-$n_f$ limit/beyond the contribution of the subset of bubble diagrams. First of all, we believe that the convergence of the $\overline{\rm MS}$ RG-functions in the large-$n_f$ limit is in some sense expected. Indeed, we know that gauge theories for $d<4$ 
at large $n_f$ flow to a CFT in the IR. In this case $\gamma$ and $\beta^\prime\equiv \partial_{\lambda_f}\beta$ at the fixed point are physical observables that at large $n_f$ depend only on $d$ and are analytic around $d=4$.
We can then use the $\epsilon$-expansion in $d=4-2\epsilon$ to turn the function analytic around $d=4$ into the same function analytic around $\lambda_f=0$. 
By analytic continuation in $d$, the same expression should hold in $d=4$, giving us the large-$n_f$ form of the $\overline{\rm MS}$ $\beta$ and $\gamma$. In order to make this point clear, we review  in appendix \ref{app:largenf} how $\gamma$ (and similarly $\beta$) can in fact be computed in this way.

 \begin{figure}[t]		
\centering			
\includegraphics[scale=.5]{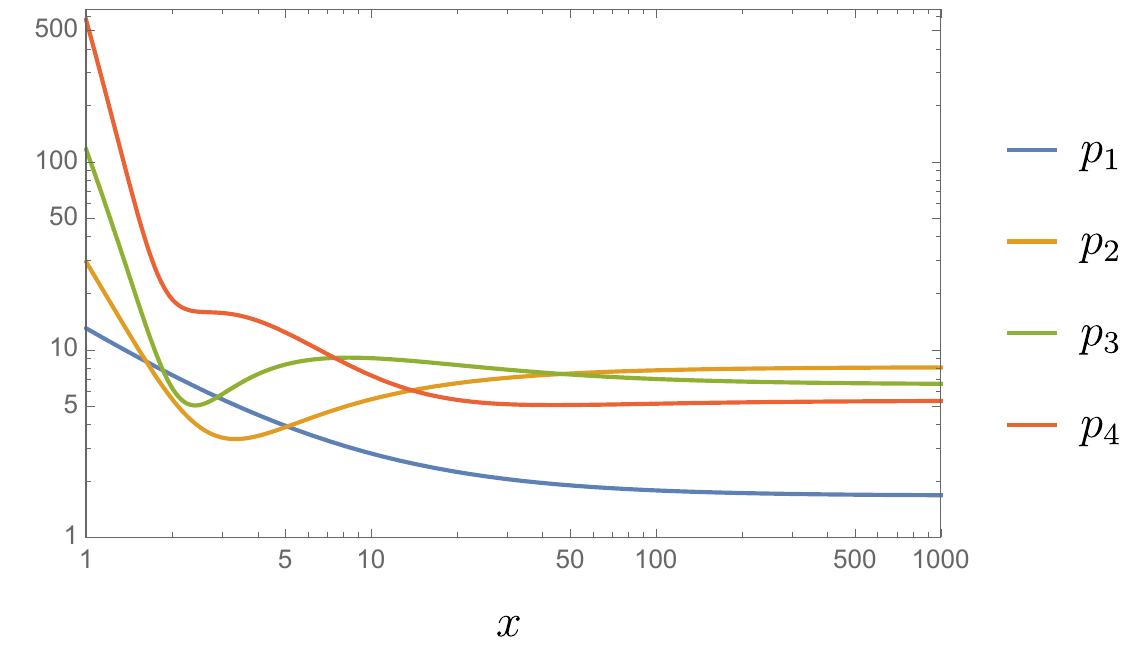} 
	\caption{The coefficients of the perturbative expansion of the $\overline{{\rm MS}}$ $\beta$-function in $\lambda_f = n_f a$ in the Veneziano limit, see eq. \eqref{eq:betaVenlargenf}. The plot shows $6-p_i(1/x)$ as a function of $x$ in a log-log scale, the offset was chosen so that the functions stay positive in the full range of $x$. We see that for large $x$ the perturbative expansion appears to be much better behaved than for $x\sim 1$. }
	\label{fig:ci}
\end{figure}

The second observation is that even though the bubble diagrams give a finite contribution, the series might still be divergent because of the contribution of other diagrams. As an indication in this direction, we can check that the perturbative expansion in the Veneziano limit for generic $x$ is worse behaved than in the large-$n_f$ (i.e. large-$x$) limit, by simply looking at the coefficients that have been computed, i.e. up to five-loop order. To this end, we write the full $\beta$ in the Veneziano limit as a function of $\lambda_f$  
\begin{equation}
\beta(\lambda_f) = \frac{2}{3} \lambda_f^2 +\frac{1}{x}\left(-\frac{11}{3}\lambda_f^2 + \sum_{i=1}^{\infty} p_i\left(\frac{1}{x}\right) \lambda_f^{i+2} \right)~.\label{eq:betaVenlargenf}
\end{equation}
The $p_i$ are polynomials of degree $i$, hence in the limit $x\to\infty$ they approach constant values $p_i(0)$. These constants match the $\mathcal{O}(n_c)$ part of the Taylor coefficients of the ${\cal O}(1/n_f)$ terms of the large-$n_f$ $\beta$-function (i.e. the analytic function $\beta^{(1)}$ in eq. \eqref{eq:betagrac}) and therefore they give a convergent series. To test the growth of these coefficients away from the large-$n_f$ limit, we use the five-loop perturbative result to obtain the $p_i$'s up to $i=4$. In fig. \ref{fig:ci} we show that indeed the coefficients $p_{1,2,3,4}\left(\frac{1}{x}\right)$ are much larger, and growing with increasing loop order, for $x\sim1$ than they are in the limit $x\to \infty$. We have checked that the behavior is analogous for $\gamma$ and a similar behavior is observed in QCD at finite $n_c$ as a function of $n_f$. We take this observation as an indication that away from the large-$n_f$ limit the perturbative expansion of $\overline{{\rm MS}}$ RG functions is divergent asymptotic.

The third observation is that it is natural to expect nonperturbative corrections that invalidate the results of perturbation theory --independently of the nature of the series-- when the renormalization scale is of the order of the dynamically generated scale $\mu\lesssim \Lambda_{\text{QCD}}$. 
$\overline{{\rm MS}}$ is a mass-independent scheme and to all orders in perturbation theory no mass scale enters in the RG functions, which are given by a power series in the coupling $a$. 
Non-perturbatively, however, we expect contributions to the RG functions given by a certain power of the dimensionless ratio $\Lambda_{\text{QCD}}/\mu$, accompanied by its own series of perturbative corrections, where
\begin{equation}
\Lambda_{\text{QCD}} \approx \mu \, e^{-\frac{1}{2\beta_0 a}}
\end{equation}
is the dynamically generated scale.  More precisely, we conjecture that the $\beta$ function (and similarly $\gamma$) is given by a trans-series 
\begin{equation}
\beta(a) \sim -\sum_{n=0}^\infty \beta_n a^{n+2} - \sum_{m=2}^\infty e^{-\frac{m}{\beta_0 a}}\sum_{n=0}^\infty \beta_{m,n} a^{n+2} + \ldots ~.\label{eq:transseries}
\end{equation}
We omitted for simplicity terms involving powers of $\log a$ that could be present as well.
The $\ldots$ represent other possibly present non-perturbative contributions, such as those given by instanton anti-instanton configurations. These are larger than the non-perturbative corrections in eq. \eqref{eq:transseries} only close to the upper edge of the conformal window at finite $n_c$, where all non-perturbative corrections are anyhow negligible, so they can be neglected altogether. It is interesting to ask what is the physical meaning of such corrections to the $\beta$ function, also because this will give us a handle on the estimation of the error. A natural guess is that these contributions are associated to higher dimensional operators. An irrelevant operator in the UV with dimension $4+k$ has an associated coupling $h$ of dimension $-k$. By dimensional analysis this can only appear in the $\beta$ function in the combination $h\Lambda_{\text{QCD}}^k$, which in terms of the dimensionless $\overline{{\rm MS}}$ coupling $\hat{h}= h \mu^k$ can be rewritten as
\begin{equation}
\delta\beta\sim  \hat{h} \left(\frac{\Lambda_{\text{QCD}}}{\mu}\right)^k =  \hat{h} \,e^{-\frac{k}{2\beta_0 a}}~.
\label{eq:betaNP}
\end{equation}
It has been conjectured that at the lower edge of the conformal window conformality is lost by annihilation of two fixed points \cite{Kaplan:2009kr}.\footnote{This scenario has been rigorously established in a weakly coupled, UV complete and unitary theory in \cite{Benini:2019dfy}.}
At large $n_c$ this operator is expected to be a double-trace operator \cite{Gubser:2002vv,Kaplan:2009kr,Gorbenko:2018ncu}, and four-fermion operators are good candidates. We expect this scenario remains qualitatively valid also at finite $n_c$ and hence as we approach the lower edge of the conformal window, some (or more) dimension 6 operator(s) (with a sizable four-fermion operator component) should become effectively marginal along the RG.
This argument motivates us to conjecture that the leading non-perturbative correction in eq. \eqref{eq:transseries} arise from $m=2$ and is precisely due to such dimension 6 operators.\footnote{This is the approach taken in (uncontrolled) truncations of exact RG flow equations, where the $m=2$, $n=0$ non-perturbative correction in eq.(\ref{eq:transseries}) is included and somehow estimated, see e.g. \cite{Kusafuka:2011fd}.}
In particular, no contribution with $m=1$ should arise in eq. \eqref{eq:transseries}. 
The imprints of such non-perturbative corrections might be visible in the perturbative expansion as IR renormalon singularities. This is however not necessarily the case. 
A perturbative series might be free of ambiguities, without renormalon or other kind of singularities, and hence technically Borel resummable, and still fail to reproduce the exact result, because non-perturbative effects are not captured. 
It might also be possible that (some of) the non-perturbative contributions in eq.~\eqref{eq:transseries} are not altogether visible in ${\rm \overline{MS}}$ if such scheme remains mass-independent beyond perturbation theory. Independently of these interpretations, however, we expect that non-perturbative corrections affect the RG flow and eq.\eqref{eq:betaNP} with $k=2$ gives the order of magnitude of their leading effect.

Summarizing, we will assume that the perturbative series in the coupling of the $\overline{{\rm MS}}$ $\beta$-function is divergent asymptotic and non-Borel resummable. 

\subsection{Banks-Zaks Conformal Expansion}
\label{subsec:BZCE}

Interacting CFTs are generally strongly coupled and not accessible in perturbation theory. A non-trivial zero of a $\beta$-function requires a cancellation among different orders in the perturbative expansion,
which is in manifest contradiction with the fact that a term of order $n+1$ should be parametrically smaller than one of order $n$. A way out is to assume that the lowest order coefficient $\beta_0$ is accidentally small
and to violate ``once" the perturbative rules. This is at the base of the Caswell-Banks-Zaks trick \cite{Caswell:1974gg,Banks:1981nn}. If $\beta_0>0$ and $\beta_1<0$ (recall our definition of $\beta$ in eq.(\ref{eq:betaDef})), we get a perturbative IR stable fixed point and the ordinary perturbative series in the coupling constant turns into a series expansion in $\beta_0$. At fixed $n_c$, or in the Veneziano limit, this is equivalent to an expansion in the parameter  
\be
\epsilon\equiv \frac{2}{321} (n_f^+-n_f)   \quad ({\rm QCD})\,, \quad \quad \epsilon\equiv \frac{4}{75} (x^+-x)   \quad  {\rm (Veneziano)}\,,
\label{eq:epsDef}
\ee 
where
\be
n_f^+ = \frac{33}2 \quad ({\rm QCD})\,, \quad  \quad \quad x^+=\frac{11}2 \quad \quad  \quad{\rm (Veneziano)}
\ee
represent the upper edge of the conformal window in the two cases.
The fixed point equations $\beta(a^*)=0$, $\beta(\lambda^*)=0$ can be solved perturbatively in $\epsilon$ and give
\be
a^*= \sum_{n=1}^\infty b_n \epsilon^n  \quad ({\rm QCD})\,, \quad \quad \lambda^*= \sum_{n=1}^\infty b_n^{{\rm V}} \epsilon^n   \quad  {\rm (Veneziano)}\,.
\label{eq:astar}
\ee
The factors $2/321$ and $4/75$ in eq.\eqref{eq:epsDef} are such that the first coefficients in the series \eqref{eq:astar} equal $b_1=b_1^{\rm V}=1$.
Using eq. \eqref{eq:astar}, the coupling constant expansion of any quantity turns into an $\epsilon$-expansion for that quantity evaluated at the fixed point.  In what follows we will call this expansion in $\epsilon$ ``conformal expansion''. In particular, for $\gamma$ and $\gamma_g$ we have 
\begin{align}
\gamma^* \equiv \gamma(a^*)  & = \sum_{n=1}^\infty g_n \epsilon^n  \quad ({\rm QCD})\,, \quad \quad \;\;
\gamma^* \equiv \gamma(\lambda^*)  = \sum_{n=1}^\infty g_n^{{\rm V}} \epsilon^n   \quad  {\rm (Veneziano)}\,,  \nn \\
\gamma_g^* \equiv  2\beta^\prime(a^*)  & = \sum_{n=1}^\infty b_n^\prime \epsilon^{n+1}  \;\; ({\rm QCD})\,, \quad \quad 
\gamma_g^* \equiv 2\beta^\prime(\lambda^*) = \sum_{n=1}^\infty b_n^{\prime {\rm V}} \epsilon^{n+1}   \;\;  {\rm (Veneziano)}\,.
\label{eq:gammastar}
\end{align}
Note that if we know $\beta$ up to loop order $L$, we can determine the conformal coefficients $b_n$, $b_n^{\rm V}$, $g_n$, $g_n^{\rm V}$, $b_n^\prime$ and $b_n^{\prime {\rm V}}$ up to order $n=L-1$.
 
From the knowledge of the first few available terms, conformal expansions appear to be better behaved than the corresponding ordinary coupling constant expansions, see e.g.\cite{Brodsky:2000cr}.\footnote{Very recently the conformal expansion  has been used to study the lower edge of the conformal window  \cite{Kim:2020yvr}. Assuming that $\gamma$ equals exactly one at the lower edge, it has been found that $n_f^*\approx 10$ in QCD. We do not think that the lower edge of the conformal window can be found in this way. First of all, as we mentioned, when $\gamma\sim-1 $ one (or more) four-fermion operator(s) is (are)  expected to be marginal and the RG flow should involve the corresponding coupling constant(s). Second, the condition $\gamma=-1$ applies possibly at large $n_c$ and corrections are expected at finite $n_c$. Third, it is not clear how the loss of conformality is explained in the approach of \cite{Kim:2020yvr}, where conformality would seem to extend below $n_f^*$.\label{footnoteConfExp}} Since in the conformal expansion we never flow away from the fixed point, and hence no dynamical scale is generated, IR renormalons are not expected to appear \cite{Gardi:2001wg}.\footnote{This is in agreement with the following fact. When an OPE approach is available, renormalons are associated to non-perturbative condensates. On the other hand, in a CFT all operators, except the identity, mush have a vanishing one-point function.}  The large order behaviour of the conformal series for $a^*$, $\gamma^*$ and $\gamma_g^*$ is unknown.\footnote{The conformal series for $a^*$ depends on the scheme, because the definition of the coupling does, and we always mean the coupling in the $\overline{{\rm MS}}$ scheme. On the other hand, since $\gamma^*$ and $\gamma_g^*$ are observables, their series expansion in $\epsilon$ is scheme-independent (i.e. each coefficient is scheme-independent).} Like for the analogue coupling constant expansions, it is not even known whether they are convergent or divergent asymptotic. If the series in the coupling is divergent asymptotic, as we argued in section \ref{sec:gce} the conformal series is necessarily divergent as well.
Indeed, if the series expansion \eqref{eq:betaDef} of $\beta(a)$ is divergent, so is also the series of the function $\epsilon(a)$ obtained by solving $\beta(a) = 0$, because the series for $\epsilon(a)$ is simply obtained from that of $\beta(a)$ dividing by a constant times $a^2$. Now suppose by contradiction that the series \eqref{eq:astar} for the inverse function $a(\epsilon)$ is instead convergent, namely that at the point $\epsilon=0$ the function is analytic. Since $a^\prime(0)\neq 0$, we could apply Lagrange inversion theorem to prove that the function $\epsilon(a)$ is also analytic at the origin, contradicting the initial assumption that the series (\ref{eq:betaDef}) for $\beta(a)$ is divergent. Due to the factorial growth of asymptotic series, the inverse series $a(\epsilon)$ is not only expected to be asymptotic, but also to have the same leading large order behaviour, modulo an overall factor.\footnote{In fact, it has been argued in \cite{Brezin:1976vw}, inverting asymptotic series, that the most notable and studied conformal expansion, the $\epsilon$-expansion in quartic scalar theories  \cite{Wilson:1971dc}, is divergent asymptotic. By now the numerical evidence of the asymptotic nature of the $\phi^4$ $\epsilon$-expansion is overwhelming, see e.g. \cite{Kompaniets:2017yct}.} We will then assume in what follows that the conformal series (\ref{eq:astar}) and (\ref{eq:gammastar}) are all divergent asymptotic, consistently with our assumption about the coupling expansion. Even if IR renormalons might not occur,  the series will be non-Borel resummable because of the same non-perturbative effects not captured in the coupling expansion.

Independently of the nature of the perturbative series, one should be careful in using the conformal expansion because the loss of conformality is not directly visible and one could erroneously conclude that the conformal window extends below $n_f^*$. This erroneous result is due to the fact that non-perturbative corrections become sizable for $n_f \sim n_f^*$ and the analysis based on perturbation theory breaks down. As explained in the previous subsection, we expect this to happen when some operators, that have dimension 6 in the unperturbed theory, approach marginality.  

\section{Borel Resumming a Non-Borel Resummable Series: What For?}
\label{sec:NBS}

As well-known, divergent asymptotic series can provide at best an approximate description, with an accuracy that depends on the behaviour 
of their series coefficients. 
Denoting by 
\be
\sum_{n=0}^\infty F_n a^n
\label{eq:Fa}
\ee
an asymptotic perturbative expansion in a coupling $a$ of the observable $F(a)$, if 
\be
F_n \sim n! \,{\rm Re}\left[t_0^{-n}\right] \,,\quad \quad n\gg 1\,,
\label{eq:largeL}
\ee
for some complex parameter $t_0$, the best accuracy is obtained by keeping $N_{{\rm Opt}}(a) \approx |t_0|/a$ terms. We refer to this as the ``optimal truncation''
of the perturbative series. The difference $\Delta$ between the exact value of the function and the one as determined from its asymptotic series at the point $a$ 
in optimal truncation is given by 
\be
\Delta(a)  \sim e^{-\frac{|t_0|}{a}}\,.
\label{eq:Delta}
\ee
Keeping more than $N_{{\rm Opt}}(a)$ terms in the asymptotic series would lead to larger discrepancies. 

\begin{table}[t]
\centering
{\renewcommand{\arraystretch}{1.4}%
\begin{tabular}{| c| c | c | c|c|c| c|}
\hline
 Loop Order    &    
 $n_f=12$      &    $n_f=13$    &    $n_f=14$     &     $n_f=15$  &     $n_f=16$  \\
\hline
 2    &  $6\cdot 10^{-2}$     &  $3.7\cdot 10^{-2}$     &  $2.2\cdot 10^{-2}$     &   $1.1\cdot 10^{-2}$   &  $3.3\cdot 10^{-3}$  \\
 3    &   $3.5\cdot 10^{-2}$  &   $2.5\cdot 10^{-2}$    &  $1.7\cdot 10^{-2}$ &  $9.8\cdot 10^{-3}$  & $3.2\cdot 10^{-3}$  \\ 
 4    &    $3.5\cdot 10^{-2}$    &    $2.7\cdot 10^{-2}$    &   $1.8\cdot 10^{-2}$  &   $1.0\cdot 10^{-2}$  & $3.2\cdot 10^{-3}$ \\ 
 5    &     ${{\color{red}-5.5\cdot 10^{-6}}}$   &     ${{\color{red}3.2\cdot 10^{-2}}}$    & $1.9\cdot 10^{-2}$     &     $1.0\cdot 10^{-2}$  & $3.2\cdot 10^{-3}$  \\ \hline
\end{tabular} }
\caption{\small Approximate values of the QCD Caswell-Banks-Zaks fixed point coupling $a^*$  as a function of $n_f$ obtained using different loop orders. The red color indicates 
values that should be taken with care, because of a possible breakdown of perturbation theory.}
\label{table:PertBeta}
\end{table}

If the large order behaviour of a series is unknown, optimal truncation can be implemented by demanding that higher-order terms are always smaller (in absolute value) than lower-order ones. The Borel transform of the series (\ref{eq:Fa}) is defined as usual:
\begin{equation}
{\cal B}F(t)=\sum_{n=0}^\infty \frac{F_n}{n!}t^n\,,
\label{eq:BFa}
\end{equation}
and we indicate by 
\begin{equation} 
\label{eq:FB}
F_{B}(a)=\int_0^\infty  \!\!dt\, e^{-t} \,{\cal B}F(t a) 
\ee
its inverse. Independently of its Borel summability, if the coefficients $F_n$ behave as in eq. \eqref{eq:largeL}, $|t_0|$ equals the distance from the origin of the singularity of  ${\cal B}F(t)$ closest to the origin in the Borel plane $t$. The series (\ref{eq:Fa}) is ``technically" Borel resummable if and only if the function ${\cal B}F(t)$ has no singularities over the positive real $t$ axis. However it reproduces the exact result $F_{B}(a)=F(a)$ only if the function $F(a)$ 
is analytic inside a disk in the complex plane $a$ (with the origin on the boundary), with a radius that depends on $t_0$ \cite{Sokal:1980ey}. If this is not the case, the unambiguous  function $F_B(a)$ will miss some ``non-perturbative" terms.
Let us now assume the worst case scenario, namely that the series $F_n$ is not technically Borel resummable and non-perturbative corrections are missed. In this case $F_{B}(a)$ is not well-defined, because we need a contour prescription to avoid the singularities present along the $t$ axis. Let us denote by $t_1$ the distance from the origin of the closest singularity on the positive real axis.
We end up having an equivalence class of functions $\{F_{B,n}(a)\}$, one for each different prescription, and an ambiguous partial result. For simplicity let us consider the typical situation where the order of magnitude of the ambiguity is the same as the leading non-perturbative correction which is anyhow missed.\footnote{Whenever the singularities 
are due to semi-classical configurations such as instantons, one might hope to cancel the ambiguities by combining in a single trans-series the asymptotic series arising from each semi-classical configuration. In the best case scenario, one could be able in this way to recover all missing non-perturbative effects and reproduce the exact result.
This is the subject of resurgence, see \cite{Aniceto:2018bis} for a recent review.}
The difference $\Delta_B$ between the exact value of the function $F$ and any $F_B$ in the class $\{F_{B,n}(a)\}$ is roughly given by 
\be
\Delta_B(a) \sim e^{-\frac{t_1}{a}} \,.
\label{eq:DeltaB}
\ee
The error \eqref{eq:DeltaB} should be compared with the best one we can obtain in perturbation theory using optimal truncation, given in eq. \eqref{eq:Delta}. 
Crucially, the singularity at $|t_0|$ and the one at $t_1$ are generally different.
Since by definition $t_1\geq |t_0|$,  {\it Borel resumming a formally non-Borel resummable function might lead to a better accuracy in the ending result.}

After this general discussion, we can come back to the situation at hand, and see what ordinary gauge coupling perturbation theory
predicts for the zeroes of the $\overline{{\rm MS}}$ $\beta$-function.
We report in table \ref{table:PertBeta} the values $a^*$ where $\beta(a^*)$ vanishes for QCD for different values of $n_f$ and using different loop orders.
As expected, the closer we are to the upper edge $n_f^+ = 33/2$ of the QCD conformal window, the more reliable perturbation theory is. 
The five-loop term for $n_f=12$ and $n_f=13$  is larger than the lower-order (four-loop) term for the values of $a$ where
the four-(or lower-)loop $\beta$ function has zeroes. For this reason they are reported in red and should be taken with care.
Barring numerical accidents and assuming the series entered its asymptotic form, a higher-order term 
 larger in magnitude than a lower-order one in a series signals either that i) we are outside the convergence radius of a series or that ii) the series is asymptotic.
As discussed in the previous section, we will assume the case ii) in the following.
 \begin{figure}[t]		
\centering			
\includegraphics[scale=.30]{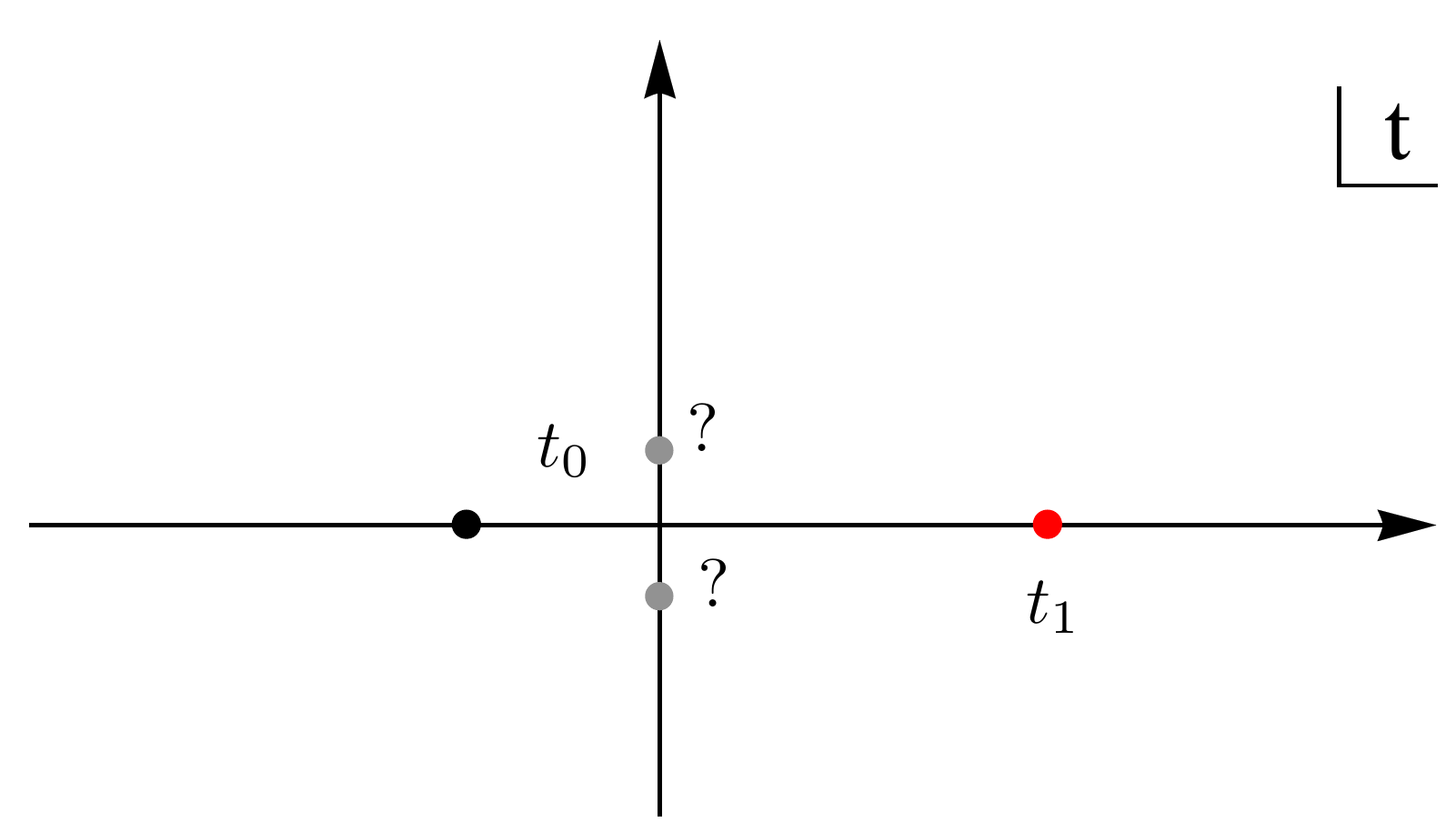} 
	\caption{Possible conjectured position of the singularities closest to the origin (black, red and grey dots) of the Borel transform of the  $\overline{{\rm MS}}$  $\beta$-function. The red singularity might be
	due to an IR renormalon or to an instanton anti-instanton configuration, the black one to a UV renormalon, the grey ones of unknown  origin.}
	\label{fig:tplanemod}
\end{figure}

The values of $t_0$ and $t_1$ are not known for $\beta(a)$. If non-perturbative corrections show up as ambiguities in the Borel resummation, as is often the case, 
based on the arguments in section \ref{sec:nature} we expect that $t_1\geq 1/\beta_0$. In particular, the singularity at $t_1=1/\beta_0$ would be
an IR renormalon related to operators that have dimension six in the UV. For an instanton anti-instanton configuration at finite $n_c$ we would have $t_1=1$, but this is always either sub-leading or negligible, as we discussed.\footnote{In the renormalon literature, irrelevant operators are typically associated to UV renormalon singularities, that appear 
in the positive real $t$ axis in a non-asymptotically free theory. They are the manifestation of the non-perturbative non-renormalizability of the theory due to Landau poles.
Although some formal similarities, the two scenarios should not be confused.}
The singularity $t_0$ might be in the negative real axis and due to a UV renormalon, though we cannot exclude the presence of other singularities off the real axis of unknown origin.\footnote{As a matter of fact, the first five loop coefficients for $\beta(a)$, both in QCD and in the Veneziano limit, alternate in signs every two. Of course we cannot draw any conclusion from the first few order terms, but such behaviour would match with a 
pair of complex conjugate leading singularities close to the imaginary axis.}  The scenario is summarized in fig.~\ref{fig:tplanemod}.
The first UV renormalon might occur at  $t_0=-1/(2\beta_0)$, in which case we would indeed have $t_1> |t_0|$.
Note that both $t_1$ and $t_0$, being proportional to $1/\beta_0$, are parametrically far away from the origin when we approach the upper edge of the conformal window.
They move towards the origin as the values of $n_f$ or $x$ decrease in going towards the lower edge of the conformal window.

The analysis above was based on the implicit assumption that we can reconstruct 
the exact Borel function. In practice we know just five perturbative terms of $\beta(a)$, so the Borel function requires a numerical reconstruction. We use Pad\'e approximants to estimate ${\cal B}F(t)$  and refer the reader to appendix \ref{app:numerics} for the details of the numerical implementation. See also appendix \ref{app:pade} for a
brief review on convergence properties of Pad\'e approximants.
The limited number of known perturbative terms makes the numerical approximation subject to an error 
that is typically larger than the one estimated to arise from the non-Borel summability of the series and given by eq. \eqref{eq:DeltaB}. 

\section{Results}
\label{sec:results}

We report in this section our numerical results for the conformal window both for QCD with $n_c=3$ and for the Veneziano limit. 
As discussed in the introduction, the more strongly coupled region remains unaccessible to us, so the conformal window  possibly extends beyond the values of $n_f$ and $x$ we can probe.

\subsection{The Veneziano Limit}

\begin{figure}[t]		
\centering			
\includegraphics[scale=.37]{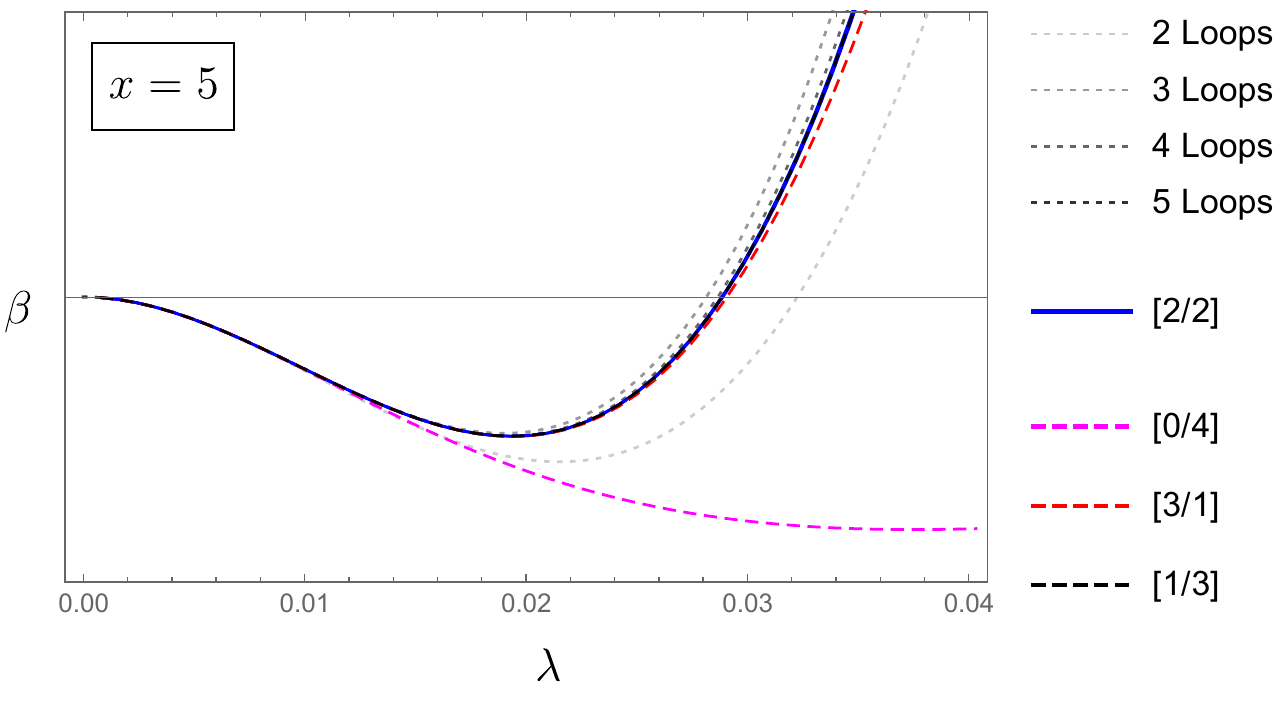} 
\includegraphics[scale=.36]{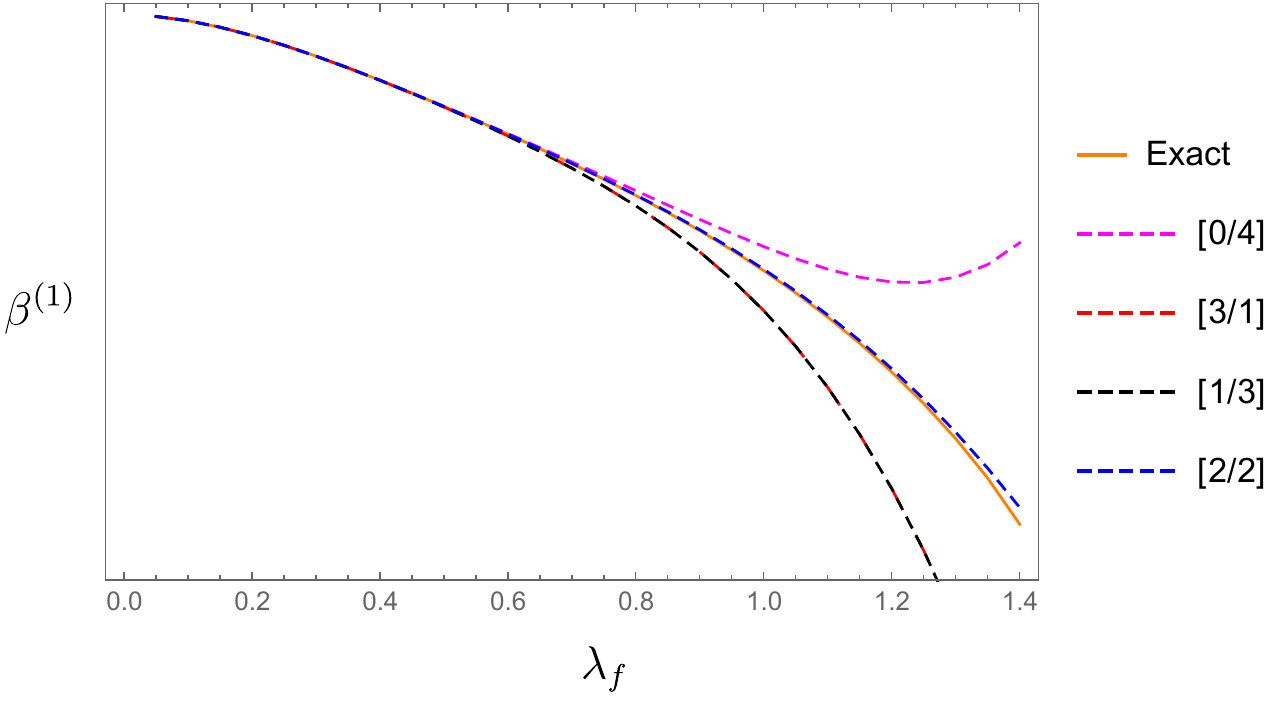} 
	\caption{Left: $\beta$-function as a function of the coupling $\lambda$ in the Veneziano limit.
	 Dotted grey lines denote perturbative results, dashed lines denote the Borel resummations using Pad\'e approximants as indicated in the legend.
		 The barely visible error band of the $[2/2]$ approximant appears here as a continuous blue line and the central values of the $[2/2]$ and $[1/3]$ approximants are overlapped. The error band corresponds to $c_{{\rm np}}=0.1$. Right: Coefficient of order $1/n_f$ of the $\beta$-function as a function of the coupling $\lambda_f$.	
		 The continuous orange line denotes the exact result obtained using large-$n_f$ methods, the dashed lines denote the Borel resummation using Pad\'e approximants as indicated in the legend.}
	\label{fig:x5_Largenf}
\end{figure}

\begin{figure}[t]		
\centering			
\includegraphics[scale=.405]{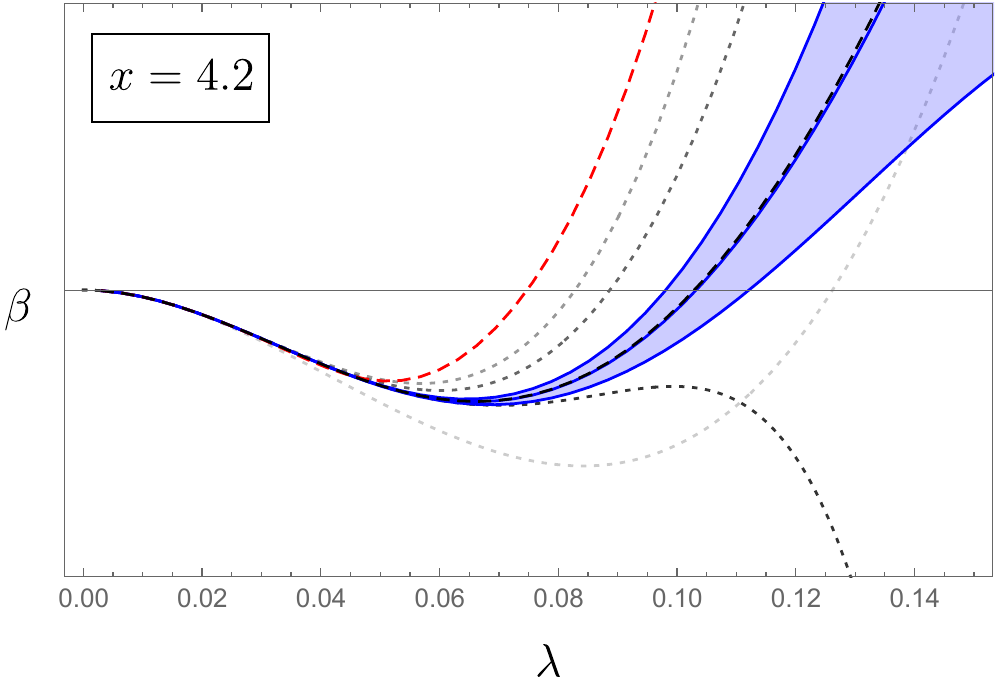} 
	\hspace*{1pt}
\includegraphics[scale=.405]{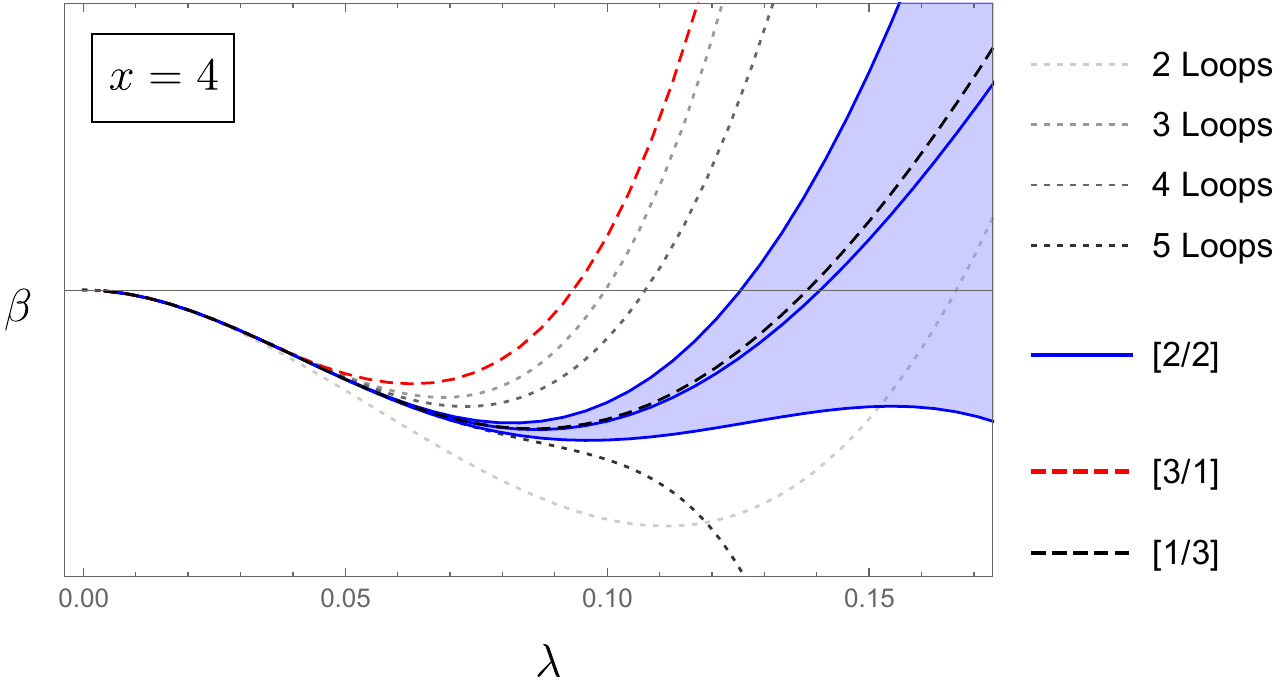} 
	\caption{$\beta$-function as a function of the coupling $\lambda$ in the Veneziano limit.
	 Dotted grey lines denote results in perturbation theory, dashed lines denote the central values of the Borel resummations using $[3/1]$ ad $[1/3]$ Pad\'e approximants as indicated in the legend.
	  The shaded area corresponds to the error associated to the $[2/2]$ approximant and its central value is given by the continuous blu line in the middle. 
	  In both panels the error band corresponds to $c_{{\rm np}}=0.1$.}
	\label{fig:x4_Largenf}
\end{figure}

In the Veneziano limit the upper edge of the conformal window is at $x^+=11/2$. 
To begin with we consider a value of $x$ relatively close to $x^+$, where the fixed-point is expected to be weakly coupled.\footnote{Strictly speaking, weakly coupled with respect to the number of available coefficient terms.} 
This is well described in perturbation theory, as shown in the left panel of fig.~\ref{fig:x5_Largenf}. 
 We note that the results for $\beta$ obtained by reconstructing the Borel function through the Pad\'e approximant $[0/4]$ are quite off compared to the other approximants and to perturbation theory, though perturbation theory 
 is supposed to be valid in this regime. For this reason we exclude from our analysis the $[0/4]$ approximant and report in what follows only the results obtained using the remaining approximants $[1/3]$, $[2/2]$ and $[3/1]$. This will also be the case for QCD.
We can actually use the exact ${\cal O}(1/n_f)$ result for $\beta$ in the  large-$n_f$ limit to test the accuracy of the different maximal Pad\'e-Borel approximants. In the Veneziano limit, large $n_f$ means large $x$ limit.
In the right panel of fig.~\ref{fig:x5_Largenf} we show the $1/x$ term in $\beta$, denoted with $\beta^{(1)}$, as a function of the large-$n_f$ coupling $\lambda_f= a n_f$. The continuous line is the exact 
result determined using large-$n_f$ methods, while the dashed lines are Pad\'e-Borel approximants constructed with the leading order large-$x$ coefficients of $\beta$ up to 5-loops.
It is evident from the figure that $[2/2]$ is the approximant that better reproduces the exact result over the whole range shown\footnote{The expansion in $\lambda_f$ of the $1/n_f$ term of $\beta$ is convergent, while we expect the series in $\lambda$  in the ordinary Veneziano limit at finite $x$ to be divergent. Moreover, at large $x$ the theory is in another phase, being obviously not UV free.
We are here assuming that the relative performance of the different approximants in the large $x$ limit applies also at finite $x$.} and thus 
we will consider this approximant as the preferred one. We will also report results for the $[3/1]$ and $[1/3]$ approximants, but only for the $[2/2]$ we will show in addition the associated (non rigorous) error, estimated as explained in appendix \ref{app:numerics}. A similar analysis applies also for $\gamma$ and $\gamma_g$, in which case the $[2/2]$ approximant is again the one that better matches the large-$n_f$ results (the evidence for $\gamma$ is not
as strong as for $\beta$ and $\gamma_g$). In the left panel of fig.~\ref{fig:x5_Largenf} and in all the other similar plots that will follow we have taken $c_{{\rm np}} =0.1$ in the non-perturbative contribution (\ref{errorNP}) to the error.

As $x$ decreases, the fixed-point occurs at larger values of the coupling $\lambda$. We report in fig.~\ref{fig:x4_Largenf} the results for $\beta(\lambda)$ at two values of $x$: $x=4.2$ and $x=4$.
The perturbative results (grey dotted lines) significantly differ from each other though they all predict the presence of a non-trivial zero, except for the 5-loop perturbative result (dark grey).
We interpret this result as a loss of reliability of the 5-loop coefficient. This interpretation is confirmed by the results obtained by Borel resumming the series.
Although the central values of the Pad\'e-Borel approximants $[1/3]$, $[2/2]$ and $[3/1]$ significantly differ from each other, they all predict a fixed point for both $x=4.2$ and $x=4$.
The error band is however different in the two cases and in particular for $x=4$ it is too large to claim the presence of conformality. 

\begin{figure}[t]		
\centering			
\includegraphics[scale=.47]{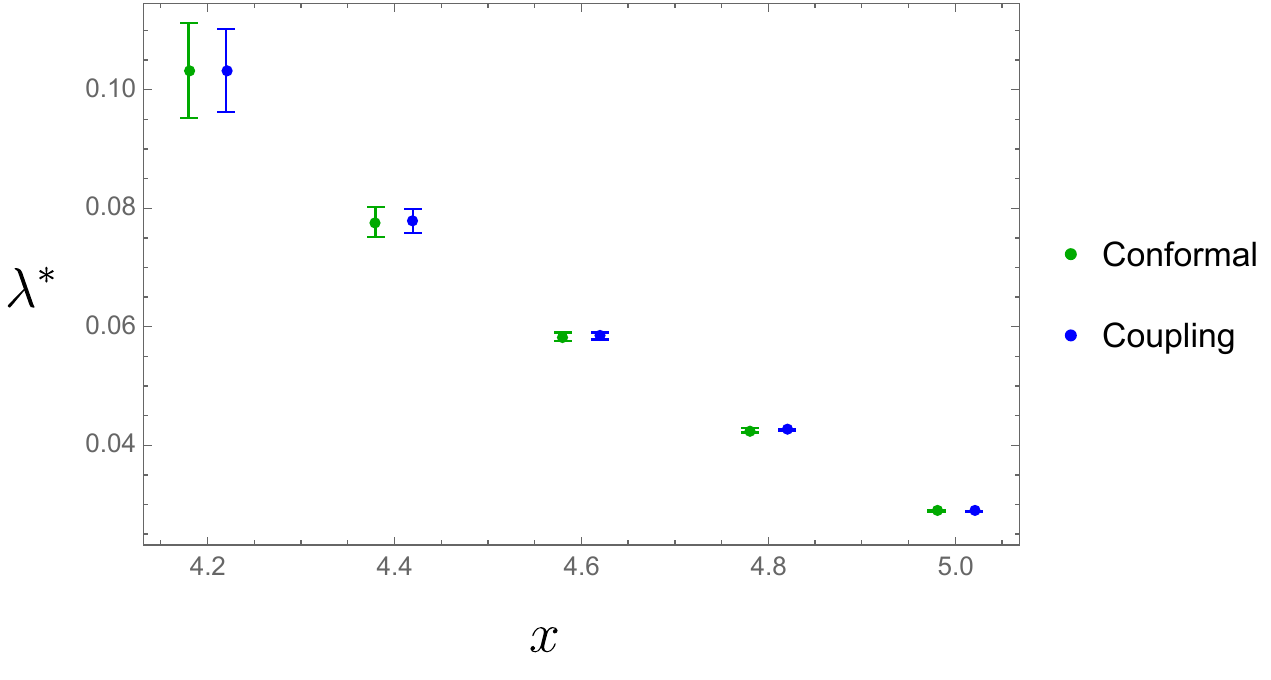} 
	\caption{The fixed point coupling as a function of $x=n_f/n_c$ in the Veneziano limit.
	The central values and error bars refer to the Pad\'e-Borel approximants $[1/2]$ in the conformal expansions and $[2/2]$ in the coupling expansion. For both ordinary coupling and conformal expansions  the error bars correspond to $c_{{\rm np}}= 0.1$. }
	\label{fig:lambdaConfVen}
\end{figure}

We can also use the conformal expansion to compute $\lambda^*$, $\gamma^*$ and $\gamma_g^*$ at the fixed point and compare the results with those obtained using ordinary perturbation theory.
The truncated available series in $\epsilon$ for both $\lambda^*$ and $\gamma^*$ start at ${\cal O}(\epsilon)$ and reach ${\cal O}(\epsilon^4)$, 
while that for $\gamma_g^*$  starts at ${\cal O}(\epsilon^2)$ and reach ${\cal O}(\epsilon^5)$. In all cases the maximal Pad\'e-Borel approximants are of order 3
and are $[2/1]$, $[1/2]$ and $[0/3]$. We can select the ``best" approximant by the following argument. For $\epsilon<0$ and small, the theory is no longer UV free but formally 
we have a conformal window in a non-unitary regime where $\lambda^*<0$.  The perturbative zero at two loops reads 
\be
\lambda^* = \frac{11-2x}{13x-34} \quad ({\rm 2 \;\; loops})\,.
\label{Venx2loops}
\ee
According to eq. \eqref{Venx2loops} the fixed point turns negative for $x>11/2$ and remains essentially constant in the limit of large $x$. 
The parametric behaviour of the $[m/n]$ Pad\'e- Borel approximant as $\epsilon$ is negative and large is of order $\epsilon^{m-n+1}$ and hence 
the approximant $[1/2]$ is the one with the correct asymptotic behaviour.
\begin{figure}[t]
\vspace{1cm}
\centering			
\includegraphics[scale=.39]{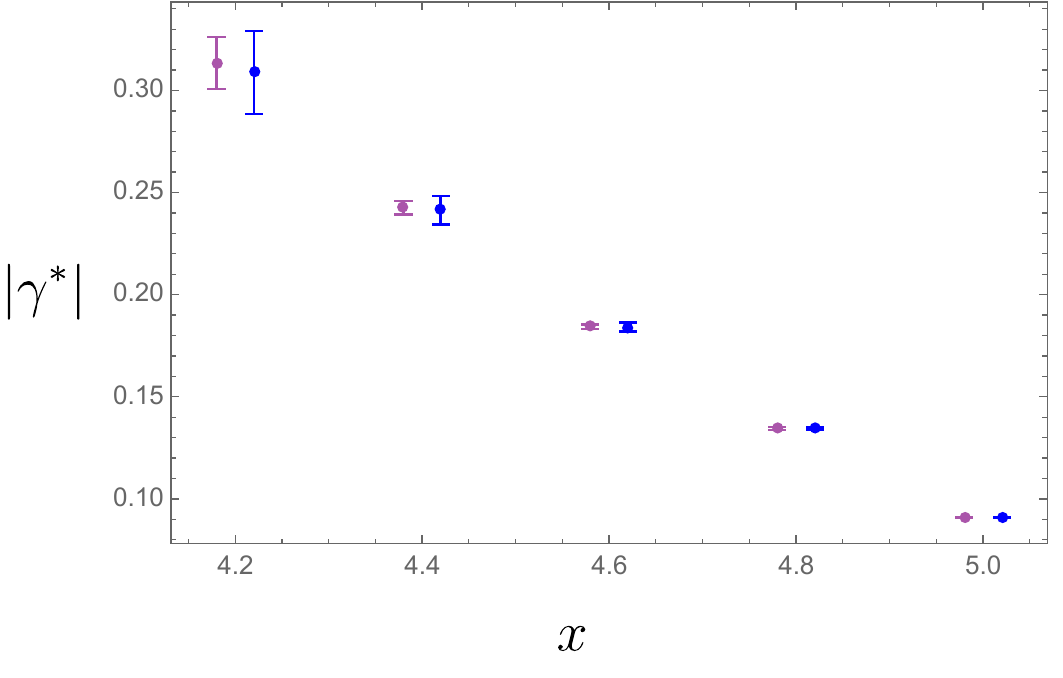} 
	\hspace*{4pt}
\includegraphics[scale=.39]{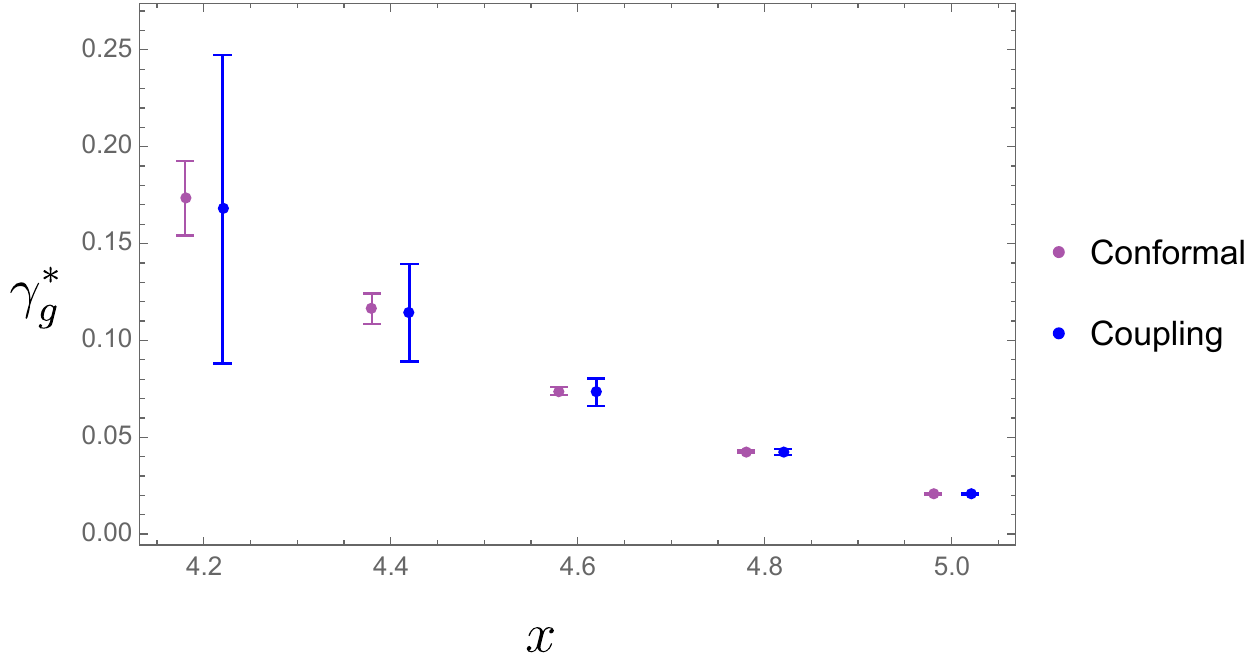} 
	\caption{Values of $|\gamma^*|$ (left) and $\gamma_g^*$ (right)  as a function of $x=n_f/n_c$ in the Veneziano limit.
	The central values and error bars refer to the Pad\'e-Borel approximants $[0/3]$ in the conformal expansions and $[2/2]$ in the coupling expansion for both $|\gamma^*|$ and $\gamma_g^*$. 
	 In both panels and for both ordinary coupling and conformal expansions  the error bars correspond to $c_{{\rm np}}= 0.1$. }
	\label{fig:gammagammagConfVen}
\end{figure}
We have verified this expectation by comparing the results for $\lambda^*$ 
obtained with the Pad\'e-Borel approximants with those obtained using the large-$n_f$ $\beta$-function for values up to $x\sim 14$.\footnote{This criterion is qualitative and not quantitative, because we are never in a controlled regime. The fixed point is under perturbative control for $x\gtrsim 11/2$, where large-$x$ results are strictly speaking not reliable.
On the other hand, at large-$x$ the perturbative fixed-point computed from the large-$x$ $\beta$-function disappears and new zeroes (possibly large-$x$ artefacts) arise close to the poles of the Gamma-functions entering in \eqref{eq:largenfbeta}. In this regime the analysis breaks down, because Pad\'e-Borel approximants can only provide an analytic continuation of the would-be perturbative fixed point. Therefore in order to compare with large-$x$ we need to assume that in the limited range  $x\lesssim 14$ the large-$x$ expansion is already good enough.}
This analysis confirms that the $[1/2]$ approximant is the best one, the $[0/3]$ having a similar performance, while the $[2/1]$ approximant has 
 the wrong asymptotic behaviour. A similar analysis applied to the Pad\'e-Borel approximants for $\gamma^*$ and $\gamma_g^*$ selects instead the $[0/3]$ as the preferred one.

We combine  in fig.~\ref{fig:lambdaConfVen} the values of $\lambda^*$ we find in both ordinary and conformal expansions using the preferred Pad\'e-Borel approximants as explained above.
The value of $\lambda^*$ is renormalization-scheme dependent, but we are always using ${\overline {{\rm MS}}}$, so a comparison is possible. It is reassuring to see that the values of $\lambda^*$ 
are in excellent agreement between themselves in the regime of interest. The main source of error arises from the numerical reconstruction of the Borel function, the theoretical one
associated to the actual non-Borel summability of the series being sub-leading, except for the $x=4.2$ case in the coupling expansion, where both error contributions are of the same order of magnitude. 

Similarly we report in fig.~\ref{fig:gammagammagConfVen} the values of $|\gamma|^*$ and $\gamma_g^*$ in both ordinary and conformal expansions using the preferred Pad\'e-Borel approximants.
Again the values of $|\gamma|^*$ and $\gamma_g^*$ are in excellent agreement between themselves in the regime of interest, with the conformal expansion being significantly more accurate, especially for $\gamma_g^*$.
The value of $\gamma^*$ is particularly useful because it can give us an indication of how far we are from the lower edge of the conformal window, assuming that $|\gamma^*|=1$ when conformality
is lost. We see from the left panel of fig.~\ref{fig:gammagammagConfVen} that at $x= 4.2$ we have $|\gamma^*|\sim 0.3$, indicating that likely the conformal window extends for values of $x<4.2$. 
Indeed, according to the conformal expansion only, the conformal window extends up to $x=4$ and below. When $x\lesssim 4$, the non-perturbative error becomes relevant and the precise
range of conformality depends on which values of $c_{{\rm np}}$ is taken in eq. \eqref{errorNP}. This is a signal that we entered a regime in which our results become unreliable. We will discuss in more detail
the impact of the choice of $c_{{\rm np}}$ on the results in the QCD case.

\begin{figure}[t!]		
\centering					
\includegraphics[scale=.55]{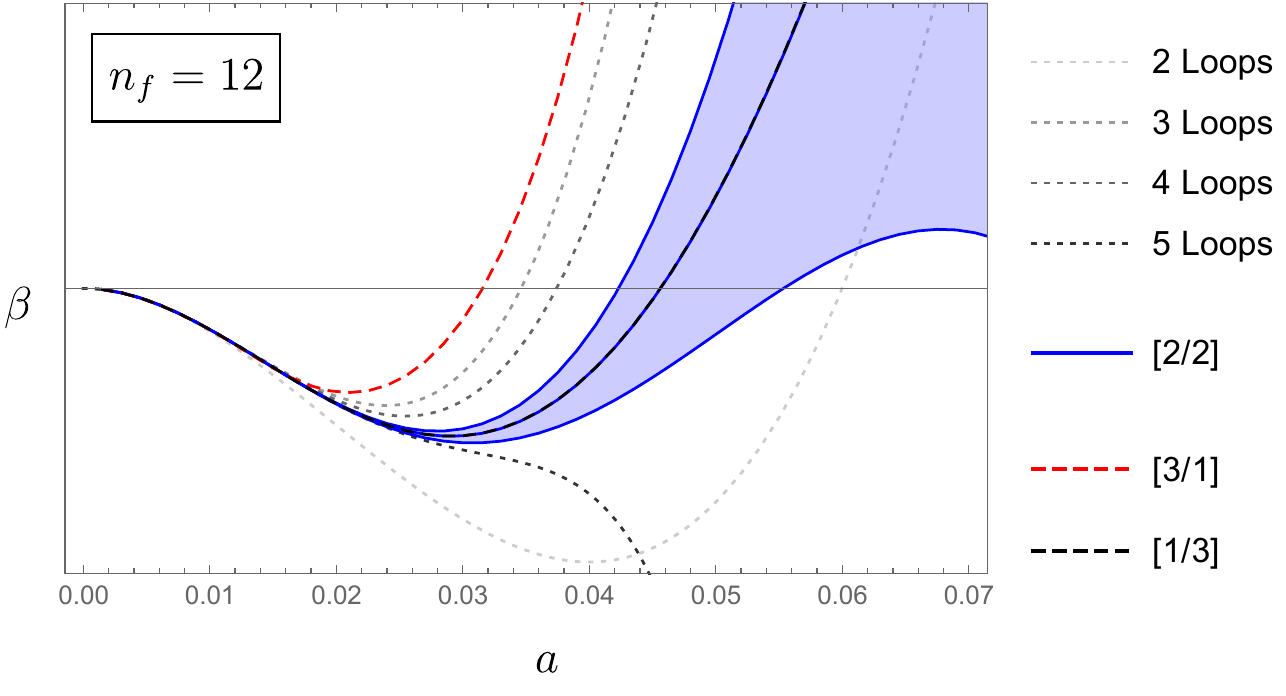} 
\centering	
\includegraphics[scale=.416]{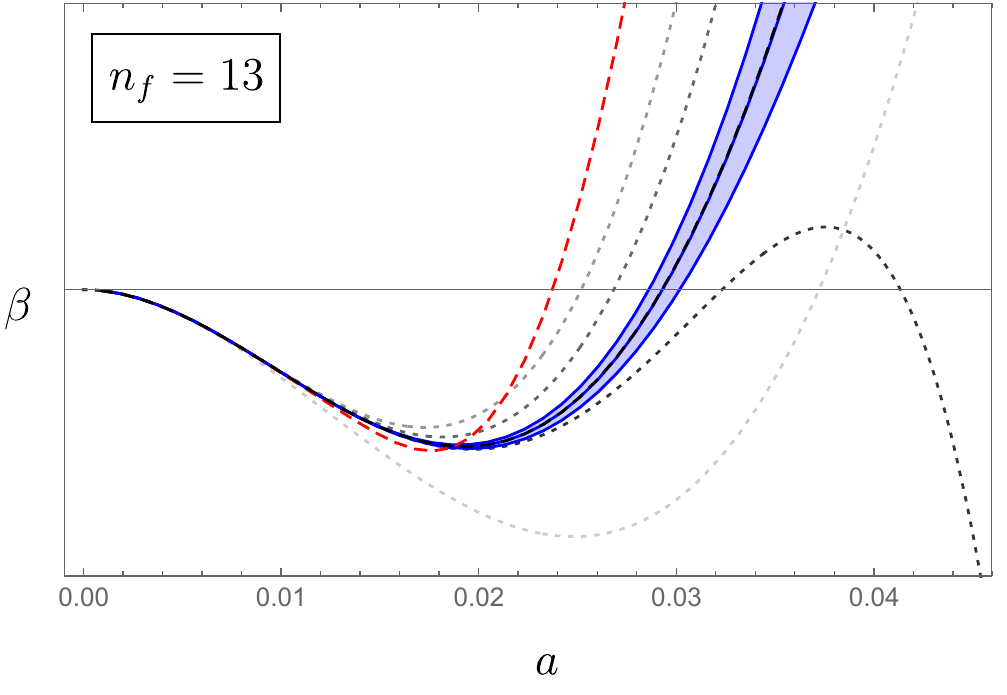} 
\includegraphics[scale=.416]{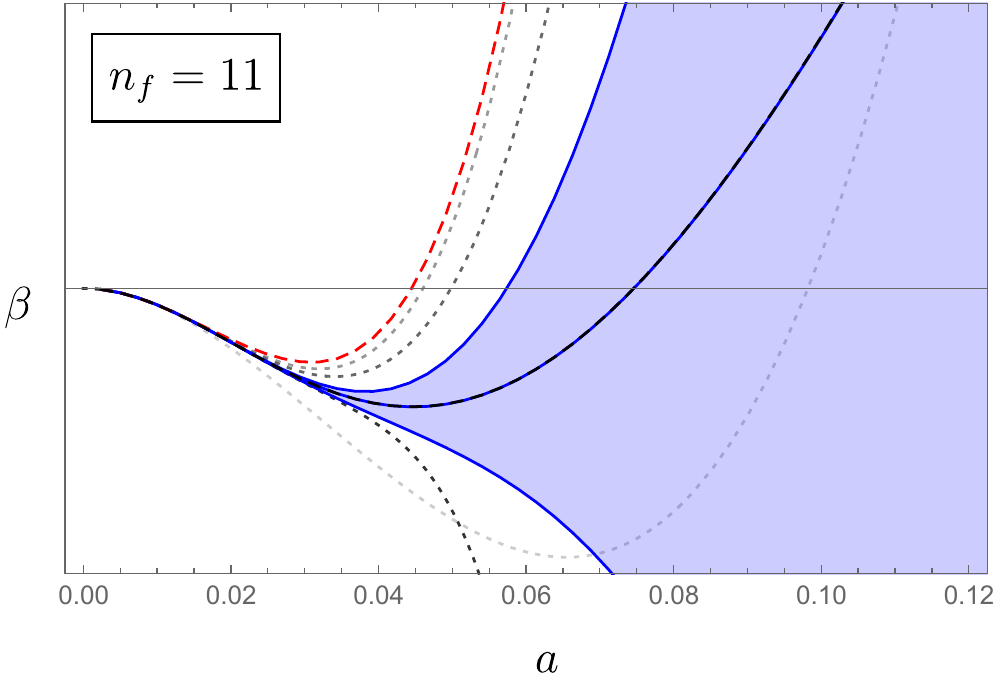} 	
	\caption{The QCD $\beta$-function as a function of the coupling $a$ for $n_f=11,12,13$ flavours.
	 Dotted grey lines correspond to perturbative results, dashed lines denote the central values of the Pad\'e-Borel approximants $[3/1]$ ad $[1/3]$ as indicated in the legend.
	  The shaded area corresponds to the error associated to the $[2/2]$ approximant and its central value is given by the continuous blu line in the middle.
			The central values of the $[2/2]$ and $[1/3]$ approximants are always overlapped. The error band corresponds to $c_{{\rm np}}=0.1$.}
	\label{fig:nf111213}
\end{figure}

Computations based on approximate Schwinger-Dyson gap equations indicate $x^* \approx 4$ \cite{Appelquist:1996dq}, which is also the value found
using truncations of exact RG flow equations \cite{Gukov:2016tnp,Kuipers:2018lux}. 
A phenomenological holographic bottom up approach gives instead $3.7 \lesssim x^* \lesssim 4.2$ \cite{Jarvinen:2011qe}.
No lattice results are available in the Veneziano limit. We do not commit ourselves with an estimate for $x^*$, which is beyond our analysis. 
However, the results shown indicate that at $x=4.2$ the theory is conformal and is probably so for values slightly below that, essentially 
in line with these earlier estimates.

\subsection{QCD: Evidence for Conformality at $n_f=12$}

In QCD the integer closest to the upper edge of the conformal window is  $n_f=16$. As can be seen from table \ref{table:PertBeta}, at $n_f=16$ 
 the fixed-point occurs for values of $a$ where perturbation theory is very accurate. Not surprisingly, the existence of an IR fixed point in this case is undisputed in the lattice community. 
As $n_f$ decreases, the fixed-point occurs at larger values of the coupling. For $n_f=15,14$ the 5-loop $\beta$-function is still reliable at these values of the coupling, as evident in table \ref{table:PertBeta} and confirmed
by Pad\'e-Borel approximants. The presence of an IR fixed point is still under perturbative control.

At $n_f\leq 13$ the 5-loop $\beta$-function coefficient is no longer reliable. We report the plots of the perturbative results and the central values of the maximal order approximants for $\beta$ as a function of $a$ for $n_f=11,12,13$
in fig.~\ref{fig:nf111213}. As in the Veneziano limit, large-$n_f$ results for both $\beta$, $\gamma$ and $\gamma_g$ select the $[2/2]$ Pad\'e-Borel approximant as the preferred one. 
We then report results for the $[3/1]$ and $[1/3]$ approximants, but only for the $[2/2]$ we also show the associated error.

\begin{figure}[t!]		
\centering			
\includegraphics[scale=.48]{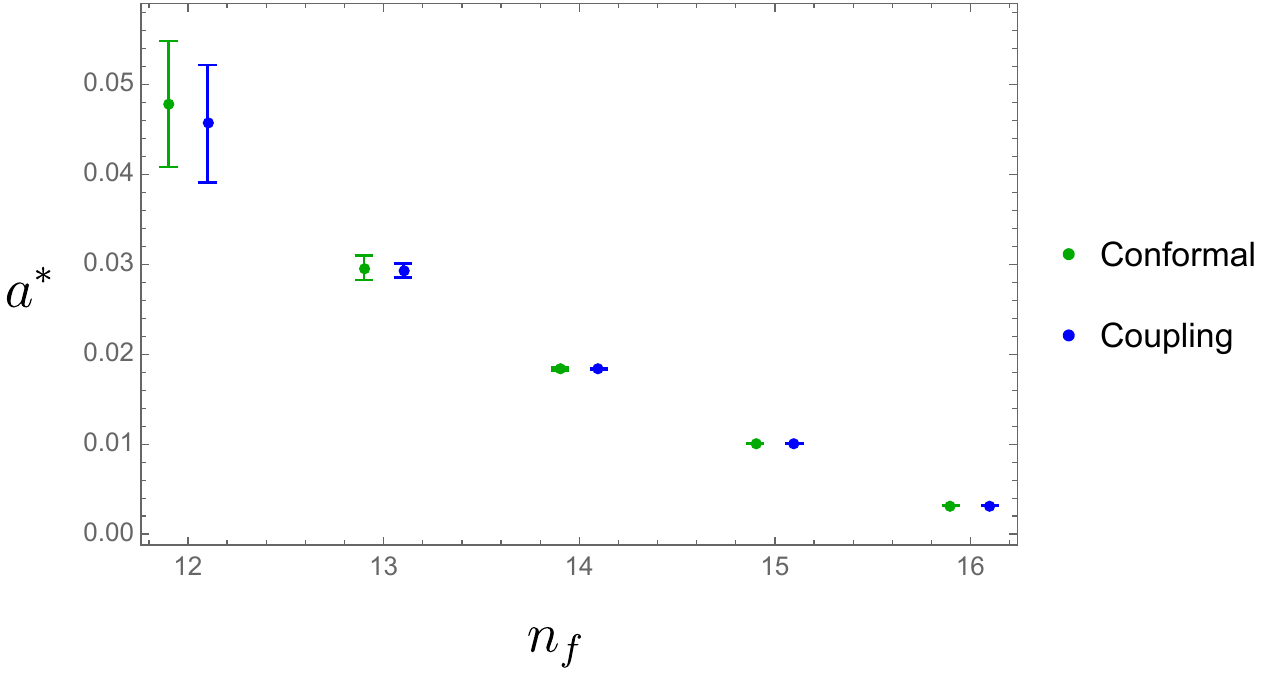} 
	\caption{The fixed point coupling as a function of $n_f$ in QCD.
The central values and error bars refer to the Pad\'e-Borel approximants $[1/2]$ in the conformal expansions and $[2/2]$ in the coupling expansion. For both ordinary coupling and conformal expansions  the error bars correspond to $c_{{\rm np}}=0.1$. }
	\label{fig:lambdaConfQcd}
\end{figure}

For $n_f=13$ the perturbative results significantly differ from each other though they all predict the presence of a non-trivial zero. The 5-loop $\beta$-function has actually two zeros.\footnote{This is in fact the case for all values of $n_f\in[13,16]$.} 
The second zero has been conjectured in \cite{Fodor:2018uih} to be related to QCD* in the scenario advocated in \cite{Kaplan:2009kr}. In contrast, we interpret this second zero 
as further evidence of the unreliability of the 5-loop result at $n_f=13$. Two-fixed points of this kind appear also in the 5-loop $\beta$-function in the Veneziano limit for $4.3\lesssim x\leq 11/2$.
In the large-$n_c$ limit the expectation is that a double-trace operator ${\cal O}$ becomes marginal at the merging point \cite{Kaplan:2009kr}. On the other hand, 
the appearance of a merging in $\beta(a)$ would imply that the single-trace operator ${\rm Tr} \,F^2$ is becoming marginal. This is in evident contradiction with our results for $\gamma_g$ in fig.~\ref{fig:gammagammagConfQcd},
which shows that this operator is irrelevant at the fixed point, having scaling dimension $4+\gamma_g$.
At finite $n_c$, the operator ${\cal O}$ can possibly contain some component of ${\rm Tr} \,F^2$, but we do not expect the merging should be visible by considering only $\beta(a)$ in perturbation theory. In fact, all the Pad\'e-Borel approximants predict a single IR-stable fixed point, as can be seen in fig.~\ref{fig:nf111213}.
We discuss further the relation between zeroes of our resummed beta functions and the QCD* scenario in the conclusions. 

We now turn to the debated case $n_f=12$.  Like at $n_f=13$, the 5-loop $\beta$-function coefficient is expected to be unreliable, even more so, due to the larger values
of the couplings explored. This expectation is confirmed by our results.
As can be seen from fig.~\ref{fig:nf111213}, all the maximal Pad\'e-Borel approximants considered show a zero of $\beta(a)$, as well as all perturbative results, but the 5-loop one. The existence of a zero for all the Pad\'e-Borel resummations within the error band indicates that $n_f = 12$ is in the conformal window.

For $n_f=11$, the value of $a^*$ as computed in perturbation theory indicates that the whole perturbative series is no longer reliable.
We see from fig.~\ref{fig:nf111213} that the qualitative picture is the same as for $n_f=12$. In particular, the central values of the Pad\'e-Borel resummations,
as well as all perturbative results but the 5-loop one, suggest a conformal behaviour.  On the other hand, the error band is too large to make any claim.\footnote{Applying Borel resummation
techniques to the perturbative $\beta$-function \cite{Antipin:2018asc} finds  $n_f^*\approx 9$ in QCD. Errors are not reported in \cite{Antipin:2018asc}. As mentioned in footnotes \ref{footnoteintro} and \ref{footnoteConfExp}, we think that the lower edge of the conformal window is not accessible with few orders in perturbation theory.}

We can also use the conformal expansion to compute $a^*$, $|\gamma^*|$ and $\gamma_g^*$ at the fixed point. Like in the Veneziano limit, an analysis of the non-unitary fixed points for $\epsilon<0$ and 
comparison with large-$n_f$ results allow us to select the Pad\'e-Borel approximant $[1/2]$ as the preferred one for the evaluation of $a^*$, and $[0/3]$ for $|\gamma^*|$ and $\gamma_g^*$.
The values of $a^*$, $|\gamma^*|$ and $\gamma_g^*$ in both ordinary and conformal expansions as a function of $n_f$ are reported respectively in figs.~\ref{fig:lambdaConfQcd} and \ref{fig:gammagammagConfQcd}.
The agreement between the two approaches is remarkable.

The value of $|\gamma^*|$ for QCD with $n_f=12$ flavours has been computed by various lattice groups over the years. We compare these results with ours in table \ref{tab:lattice}. To obtain our results we average over all the available Pad\'e approximants, weighted by the errors, with the final errors obtained combining the individual ones in quadrature (as opposed to fig.s \ref{fig:gammagammagConfVen} and \ref{fig:gammagammagConfQcd} in which we only show the result from the preferred Pad\'e approximant).
We present the results only for the conformal expansion which are more precise and less sensitive to the choice of the parameter $c_{{\rm np}}$.

Let us briefly mention the different lattice techniques that have been used to obtain the results reported in table \ref{tab:lattice}. 
The ``gradient flow'' technique of ref. \cite{Carosso:2018bmz} is based on a lattice implementation of the renormalization group. Ref. \cite{Aoki:2016yrm} measures the anomalous dimension from the scaling of the topological susceptibility with the fermion mass. Ref. \cite{Cheng:2013eu} uses the scaling of the spectral density of the massless Dirac operator. The ``finite-size scaling" technique of \cite{Lombardo:2014pda,Cheng:2013xha,Aoki:2012eq,Appelquist:2011dp} uses the dependence of correlators on the volume. As can be seen, our results are compatible with the lattice ones.
Similarly we consider $\gamma_g^*$ for QCD with $n_f=12$. In this case the only lattice result we are aware of is that of ref. \cite{Hasenfratz:2016dou}. Also for $\gamma_g$ we report the results only in the conformal expansion.  Upon performing an error-weighted average of all the available Pad\'e's and combining the errors in quadrature we get for $n_f=12$ $\gamma_g^*=0.23(6)$, in good agreement with the lattice result $\gamma_g^*=0.26(2)$ of ref. \cite{Hasenfratz:2016dou}.\footnote{Compatible values of the anomalous dimensions $\gamma^*$ and $\gamma_g^*$ for $n_f=12$ were found using the conformal expansion in \cite{Ryttov:2017lkz, Ryttov:2017kmx}, though no estimate of the error is provided there. These papers also give results up to $n_f = 9$. Again, our analysis suggests that perturbation theory is not reliable for such low values of $n_f$.}

\begin{figure}[t!]	
\centering			
\includegraphics[scale=.39]{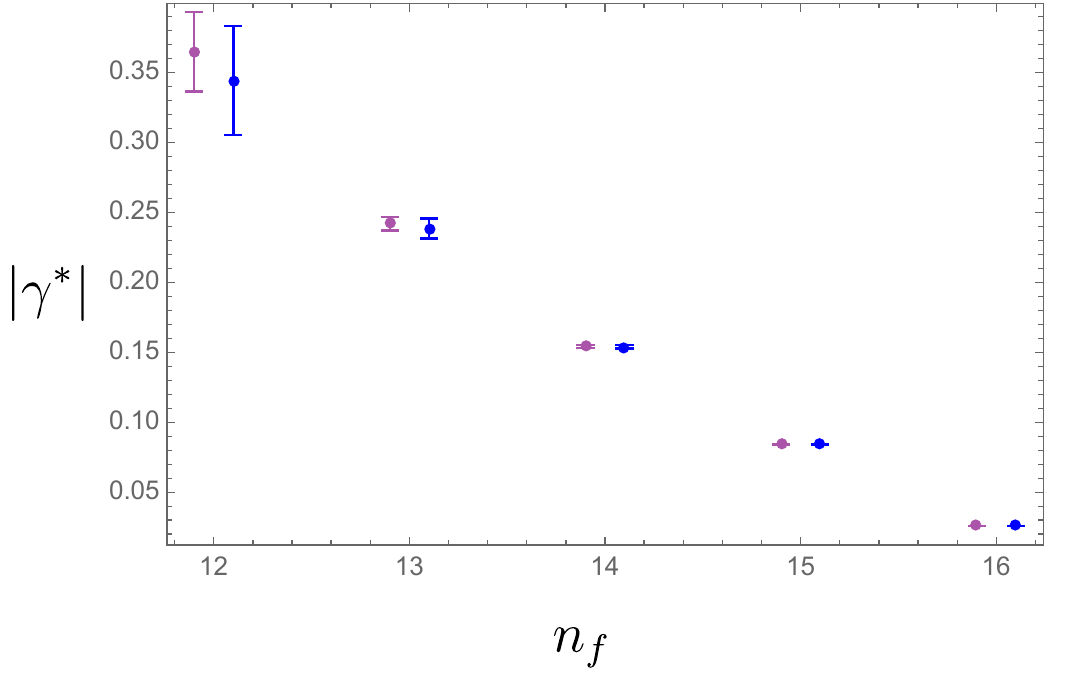} 
	\hspace*{5pt}
\includegraphics[scale=.385]{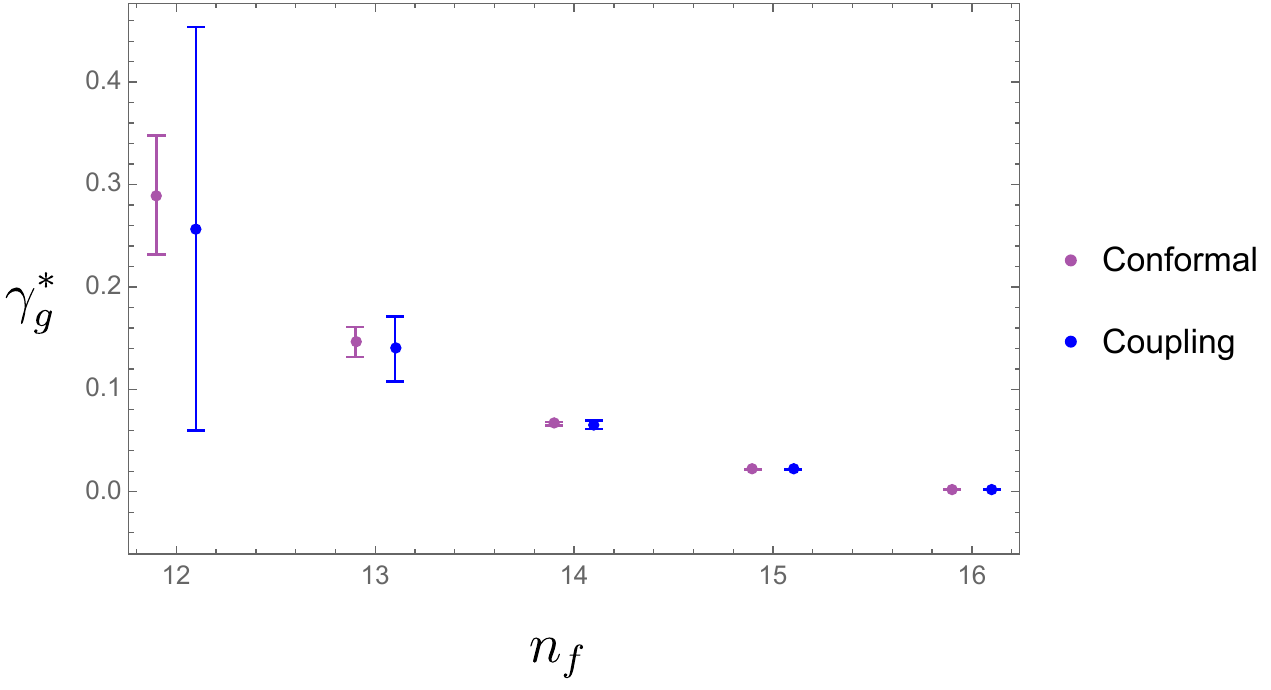} 
	\caption{
	Values of $|\gamma^*|$ (left) and $\gamma_g^*$ (right)  as a function of $n_f$ in QCD.
	The central values and error bars refer to the Pad\'e-Borel approximants $[0/3]$ in the conformal expansions and $[2/2]$ in the coupling expansion for both $|\gamma^*|$ and $\gamma_g^*$. 
	 In both panels and for both ordinary coupling and conformal expansions  the error bars correspond to $c_{{\rm np}}=0.1$. }
	\label{fig:gammagammagConfQcd}
\end{figure}

In fig.~\ref{fig:agammastarConf} we show $a^*$, $|\gamma^*|$ and $\gamma_g^*$ as a function of $n_f$ as computed using the conformal expansion with $c_{{\rm np}}=0.1$. Note that the central values of $a^*$  in both the conformal and coupling expansion are very close to each other also for $n_f=11$, i.e. the value of the central green line in the upper central panel of fig.~\ref{fig:agammastarConf} at $n_f=11$ is in good agreement with the value where the central blue line in the $n_f=11$ panel of fig.~\ref{fig:nf111213} crosses zero. This is suggestive of a conformal behaviour at $n_f=11$, but some care is needed before jumping too quickly to a conclusion.
While for $n_f\geq 12$ the choice of $c_{{\rm np}}$ is essentially irrelevant in the conformal expansion (unless one considers unreasonably large values of this parameter),
for $n_f<12$ this is no longer the case. For $n_f=12$ we have to take   $c_{{\rm np}} \sim 40$ to enlarge the error so that this is compatible with no fixed point in the conformal expansion. 
On the other hand, the fixed point for $n_f=11$ in the conformal expansion is compatible with no fixed point for $c_{{\rm np}} \sim 2$. 
We take these results as evidence that $n_f=11$ is conformal, but we do not consider it enough to make a strong claim. If we trust this evidence, we can use the resummation of the conformal expansion to obtain $|\gamma^*|=0.485(143)$ and  $\gamma_g^* = 0.36(19)$ (averaging all the available Pad\'e's weighted by their errors and combining the errors in quadrature). 
\begin{table}[t]
\centering
\begin{tabular}{c c l l}
\toprule
Ref.  &  \multirow{9}{*}{\includegraphics[scale=.519]{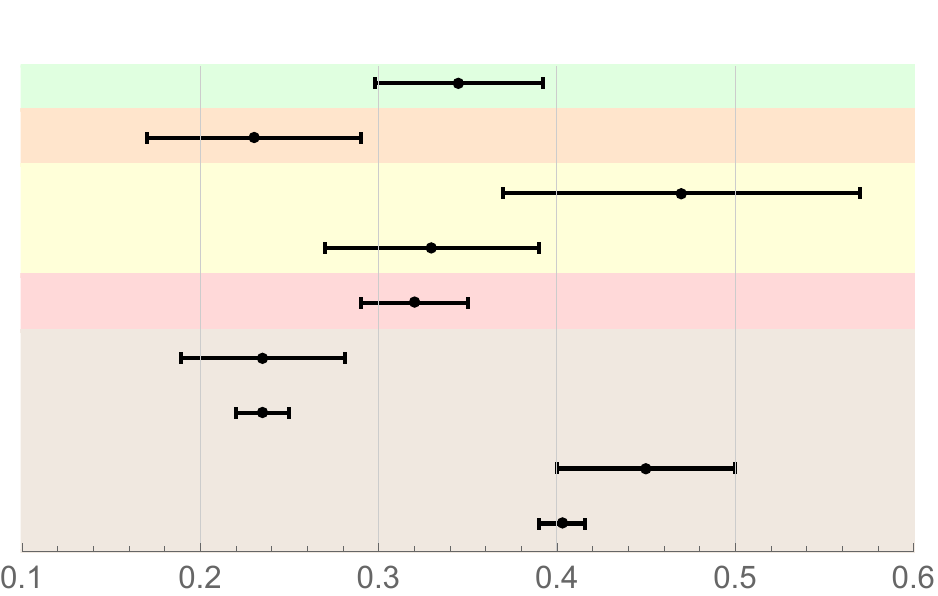}}  &  ~~~$|\gamma^*|$  &    ~~~Method     \\
\midrule \\[-16.7pt]
This work &   &  $0.345(47)$  & PB conformal \\
\cite{Carosso:2018bmz} &&  $0.23(6)$  & Gradient flow  \\
\!\!\multirow{2}{*}{\cite{Aoki:2016yrm}}  &&  $0.47(10)$  & \multirow{2}{*}{Top. susceptibility} \\
          & & $0.33(6)$      &   \\
\cite{Cheng:2013eu} &&  $0.32(3)$  & Dirac eigenmodes  \\
\cite{Lombardo:2014pda} &&  $0.235(46)$  & \multirow{4}{*}{Finite-size scaling}  \\
\cite{Cheng:2013xha} && $0.235(15)$  &   \\
\cite{Aoki:2012eq} &&  $0.45(5)$  & \\
\cite{Appelquist:2011dp} &&  $0.403(13)$  &  \\[-3pt]
\midrule \\[-8pt] \bottomrule
\end{tabular}
\caption{Comparison between the results of our Pad\'e-Borel (PB) resummation for $|\gamma^*|$ in  QCD with $n_f = 12$ using the Banks-Zaks conformal expansion and averaging over all available Pad\'e approximants, and lattice results.} 
\label{tab:lattice}
\end{table}

In contrast, the dependence on $c_{{\rm np}}$ in the coupling expansion is more severe. In fig. \ref{fig:Varyingcnp} we show how the results for $n_f=12$ change as we vary the coefficient $c_{\text{np}}$ of the non-perturbative contribution to the error. We see that increasing the coefficient from $c_\text{np} = 0.1$ to $c_\text{np} = 1$ the error band becomes too large and the result is also compatible with the absence of a fixed point. For $n_f=11$ the contribution to the error associated to the numerical reconstruction of the Borel function is so large that 
the results are compatible with no fixed point even if one takes $c_{{\rm np}}=0$. Needless to say, we do not commit ourselves with an estimate for $n_f^*$.

\begin{figure}[t]		
\centering			
\includegraphics[scale=.5]{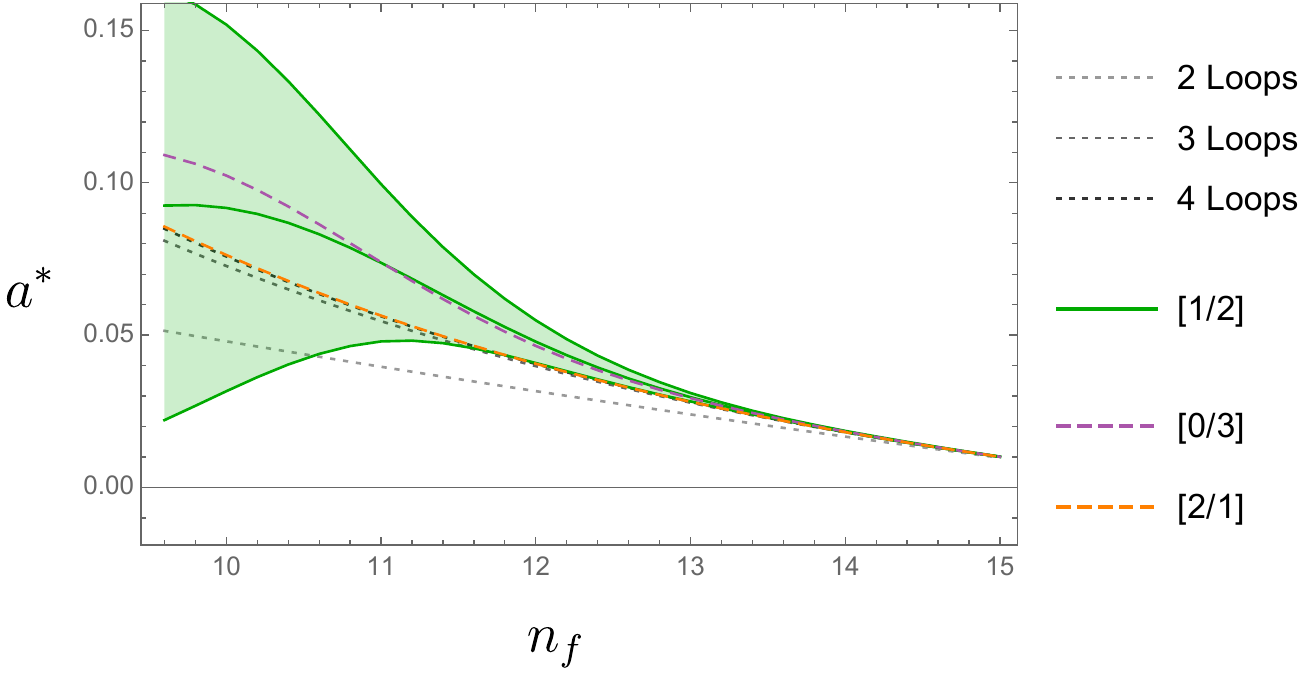} 
\centering			
\includegraphics[scale=.39]{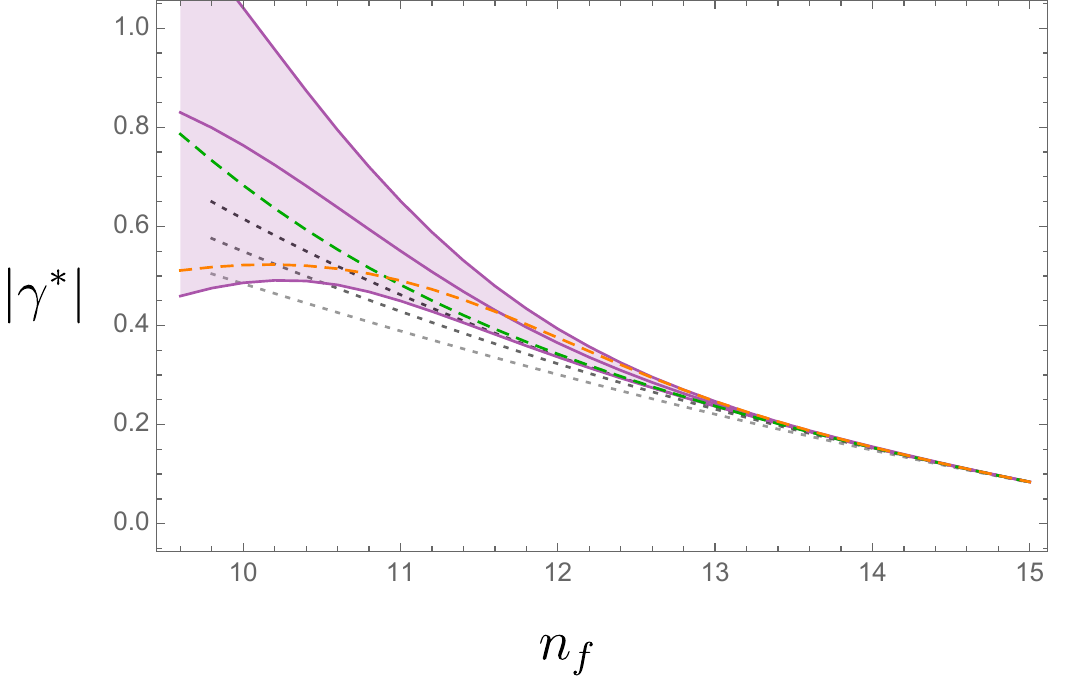} 
	\hspace*{1pt}
\includegraphics[scale=.39]{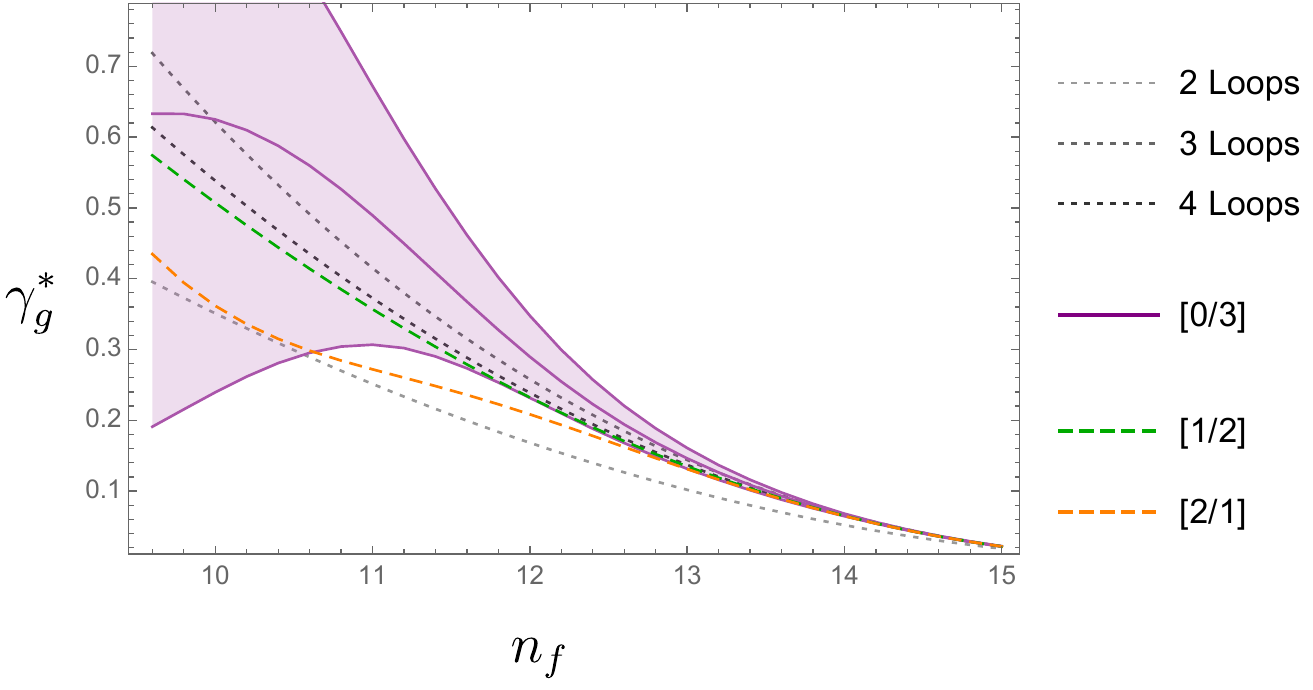} 
	\caption{
	The fixed-point coupling $a^*$ and the anomalous dimensions $|\gamma|$ and $\gamma_g$ in QCD as a function of $n_f$ as obtained from the Pad\'e-Borel resummation of the Banks-Zaks conformal expansion.
	Grey dotted lines correspond to perturbative results, dashed lines denote the central values of the Pad\'e-Borel approximants indicated in the legend.
	  The shaded area corresponds to the error associated to the selected approximant and its central value is given by the continuous line in the middle.
		In all panels the error band corresponds to $c_{{\rm np}}= 0.1$.}
	\label{fig:agammastarConf}
\end{figure}

We would finally like to conclude with a general observation about the use of Pad\'e-Borel versus simple Pad\'e approximants. 
In the former case, we would not expect a gain in considering Borel-resummation, because $\beta(a)$ would be analytic at $a=0$ and an ordinary Pad\'e approximant on $\beta(a)$ should suffice. On the contrary, for a convergent series the Borel function is analytic everywhere and expected to have an exponential behaviour at infinity, and functions of this kind are not well approximated by low-order Pad\'e approximants. We have verified that by taking ordinary Pad\'e approximants our results remain qualitatively unchanged, though Pad\'e-Borel approximants give slightly more accurate results.
This can be seen as a sort of indirect numerical evidence of the non-convergence of the $\overline{\rm MS}$ $\beta$-function. See appendix \ref{app:pade} for a more general discussion on the convergence
properties of Pad\'e approximants.

\section{Conclusions}

In this paper we applied Pad\'e-Borel resummation techniques to the RG functions of QCD, in order to test the existence of a fixed point beyond the perturbative regime. 
We considered both the ordinary expansion in the coupling and the Banks-Zaks conformal expansion. The second approach is more accurate than the former, but we showed that in the regime in which the resummation is under control they give compatible results for the critical $\overline{\rm MS}$ coupling $a^*$, and for the anomalous dimensions $\gamma^*$, $\gamma_g^*$ of respectively the fermion mass bilinear and the gauge kinetic term operator.
According to this analysis the conformal window in QCD extends at least to $n_f =12$ with weaker indications that also $n_f=11$ is included, and in the Veneziano limit the window extends at least to $x = 4.2$.

Along the way, we provided arguments that led us to conjecture that these perturbative series are divergent asymptotic, thereby supporting the need for Pad\'e-Borel techniques. 
It is important however to emphasize that the Pad\'e-Borel analysis that we used applies independently of whether the QCD $\beta$-function series is convergent or divergent asymptotic.
We also discussed the role of non-perturbative corrections, and suggested a relation between renormalon singularities in the perturbative $\beta$ function and contributions of irrelevant couplings to the running of the gauge coupling, that are invisible to all orders in perturbation theory in $\overline{\rm MS}$. Estimating the errors due to both the resummation techniques and the non-perturbative corrections was an important part of our work: while these errors are not rigorous, we believe that our estimates are reasonably conservative, and they provide a crucial sanity-check, because they forbid us to extrapolate the five-loop perturbative result to arbitrarily strong coupling. 
\begin{figure}[t!]		
\centering			
\includegraphics[scale=.48]{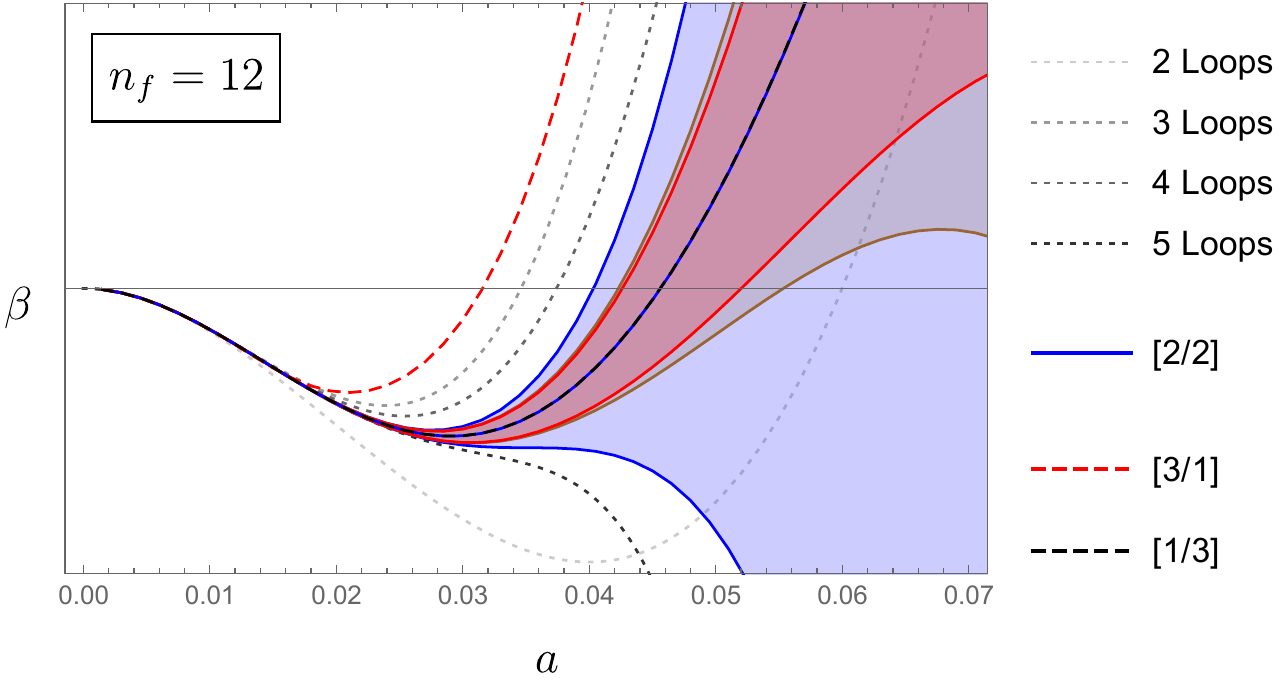} 
	\caption{The QCD $\beta$-function for $n_f=12$, for three different values of the coefficient $c_\text{np}$ in the estimate of the non-perturbative contribution to the error. The red, brown and blue bands correspond to $c_\text{np} = 0$, $c_\text{np} = 0.1$ (the same used in fig. \ref{fig:nf111213}) and $c_\text{np} = 1$, respectively.} 
	\label{fig:Varyingcnp}
\end{figure} 

We will now briefly comment about the behaviour of our resummed $\beta$ function for $n_f/x$ lower than those analyzed in the main text. In this range of $n_f/x$ the error bars become large and our central values cannot be trusted, however we observe that the central values sometimes admit a second zero, that merges and annihilate with the first non-trivial zero. For instance the merging happens for $n_f$ between 9 and 10 in QCD. It would be tempting to relate this to the QCD$^*$ mechanism that has been proposed in \cite{Kaplan:2009kr} for the end of the QCD conformal window.  However, the RG flow that connects QCD$^*$ to QCD involves four-fermion couplings: at least close to the annihilation point, the flow happens in the two-dimensional space spanned by the gauge-coupling and a four-fermion coupling that is a singlet under the global symmetry, which becomes marginal precisely when the two fixed points merge. On the other hand our analysis is completely blind to any four-fermion coupling. As we recalled above, four-fermion couplings could appear in the $\overline{\rm MS}$ $\beta$ function only via non-perturbative corrections. If we are conservative enough in estimating our non-perturbative errors, we should then see large error bars covering up the strong-coupling region in which these couplings are not negligible. This is consistent with the fact that we observe annihilation in a region where our errors are large and our resummation is not reliable. As a result, unless one finds a way to incorporate the nonperturbative corrections in the $\beta$ function, our approach is not powerful enough to study the proposed QCD$^*$ scenario for the end of the conformal window.

We regard the approach proposed here as a valid complement to non-perturbative approaches such as lattice simulations and possibly in the future also conformal bootstrap techniques. It can be useful to give an independent test when lattice results are not completely conclusive, as we have seen in the case of QCD with $n_f=12$, and also as an additional way to compute observables, as we did here for $\gamma^*$ and $\gamma_g^*$. When higher loop results will become available, they can be used to improve the precision of our numerical resummation and to lower our upper bounds on $n_f^*$ and $x^*$, at least until the point where the non-perturbative errors dominate. In the future this approach can be readily applied to more general gauge groups and matter representation, for instance to probe the conformal window in adjoint QCD, and also to other observables at the fixed point e.g. anomalous dimensions of other composite operators.

\section*{Acknowledgments}

We thank S. Cecotti and V. Gorbenko for useful discussions and J.-W. Lee for a question that allowed us to fix a mistake in a first version of the paper. We also thank M. Beneke for comments on the draft. LD and MS are partially supported by INFN Iniziativa Specifica ST\&FI.
This research project started during the ``Bootstrap 2019" workshop at Perimeter Institute. We thank the participants of the workshop for the stimulating atmosphere and the interesting discussions. Research at Perimeter Institute is supported in part by the Government of Canada through the Department of Innovation, Science and Economic Development Canada and by the Province of Ontario through the Ministry of Economic Development, Job Creation and Trade. 

\appendix

\section{Perturbative $\beta$ Function and Anomalous Dimension $\gamma$}
\label{app:betagamma}

We adopt the conventions of \cite{Herzog:2017ohr} for the $\beta$ function and its perturbative coefficients, namely 
\begin{equation}
\beta(a) = \frac 12 \frac{{\rm d }a}{{\rm d}\log\mu} = -\sum_{n=0}^\infty \beta_n \,a^{n+2}~,
\label{eq:betaDef}
\end{equation}
where we are using the variable $a \equiv \frac{g^2}{16\pi^2}$. Note that in our conventions there is a prefactor of $\frac 12$ in eq. \eqref{eq:betaDef}, and $\beta_n$ is defined with a minus sign relative to the Taylor coefficients of $\beta$. The coefficient of the lowest order $a^2$ is the one-loop coefficient $\beta_0$, and more generally $\beta_n$ is the $(n+1)$-loop coefficient. 

Similarly, our conventions for the anomalous dimension of the fermion mass operator and its perturbative coefficients are
\begin{equation}
\gamma(a) = \frac{1}{m}\frac{{\rm d} m}{{\rm d}\log \mu}=-\sum_{n=0}^{\infty} \gamma_n a^{n+1}\,,
\end{equation}
where the mass $m$ is the coefficient of $\bar{\psi} \psi$ in the Lagrangian. In this case the expansion starts at one-loop order with the power $a^1$, and similarly to the $\beta$ function $\gamma_n$ is the $(n+1)$-loop coefficient. At a fixed point $\beta(a^*)=0$, $\gamma^*\equiv \gamma(a^*)$ is related to the scaling dimension of the operator $\bar{\psi} \psi$ by 
\be 
\Delta_{\bar{\psi}\psi} = 3 + \gamma^*\,.
\label{eq:DeltaPsiDef}
\ee
The anomalous dimension of the gauge kinetic operator is
\be
\gamma_g(a) =2 \frac{\partial \beta(a)}{\partial a}\equiv 2\beta^\prime(a).
\ee 
At a fixed point $\gamma_g^*\equiv \gamma_g(a^*)$  is related to the scaling dimension $\Delta_{F^2}$ of the operator ${\rm Tr}[F^{\mu\nu}F_{\mu\nu}]$ by 
\be
\Delta_{F^2} = 4 + \gamma_g^*.
\label{eq:DeltaF2Def}
\ee

The perturbative coefficients up to five-loop order in the case of QCD read \cite{Baikov:2016tgj,Baikov:2014qja}
{\small{\begin{align}
\begin{split}
\!\!\!\!\!\!\!\!\!\!\begin{cases}
\beta_0 &=11-\dfrac{2 n_f}{3}~,~~~~~\beta_1=102-\dfrac{38 n_f}{3}~,~~~~~\beta_2=\dfrac{325 n_f^2}{54}-\dfrac{5033 n_f}{18}+\dfrac{2857}{2}~,~\vspace{0.2cm}\\
\beta_3&=\dfrac{1093 n_f^3}{729}+n_f^2 \left(\dfrac{6472 \zeta (3)}{81}+\dfrac{50065}{162}\right)+n_f \left(-\dfrac{6508 \zeta (3)}{27}-\dfrac{1078361}{162}\right)+3564 \zeta (3)+\dfrac{149753}{6}~,~\vspace{0.2cm}\\
\beta_4&=n_f^4 \left(\dfrac{1205}{2916}-\dfrac{152 \zeta (3)}{81}\right)+n_f^3 \left(-\dfrac{48722 \zeta (3)}{243}+\dfrac{460 \zeta (5)}{9}+\dfrac{809 \pi ^4}{1215}-\dfrac{630559}{5832}\right)\\
&+n_f^2 \left(\dfrac{698531 \zeta (3)}{81}-\dfrac{381760 \zeta (5)}{81}-\dfrac{5263 \pi ^4}{405}+\dfrac{25960913}{1944}\right)\\
&+n_f \left(-\dfrac{4811164 \zeta (3)}{81}+\dfrac{1358995 \zeta (5)}{27}+\dfrac{6787 \pi ^4}{108}-\dfrac{336460813}{1944}\right)\\
&+\dfrac{621885 \zeta (3)}{2}-288090 \zeta (5)+\dfrac{8157455}{16}-\dfrac{9801 \pi ^4}{20}\,,
\end{cases}
\end{split}
\end{align}}}
{\small{\begin{align}
\begin{split}
\!\!\!\!\!\!\!\!\!\!\!\!\!\!\!\!\!\!\!\!\!\!\!\!\!\!\!\!\!\!\!\!\!\!\!\!\!\!\!\!\!\!\!\!\!\!\!\begin{cases}
\gamma_0 &=8~,~~~~~\gamma_1=\dfrac{404}{3}-\dfrac{40 n_f}{9}~,~~~~~\gamma_2=-\dfrac{280 n_f^2}{81}+n_f \left(-\dfrac{320 \zeta (3)}{3}-\dfrac{4432}{27}\right)+2498~,~\vspace{0.2cm}\\
\gamma_3&=n_f^3 \left(\dfrac{128 \zeta (3)}{27}-\dfrac{664}{243}\right)+n_f^2 \left(\dfrac{1600 \zeta (3)}{9}-\dfrac{32 \pi ^4}{27}+\dfrac{10484}{243}\right)\\
&+n_f \left(-\dfrac{68384 \zeta (3)}{9}+\dfrac{36800 \zeta (5)}{9}+\dfrac{176 \pi ^4}{9}-\dfrac{183446}{27}\right)\\
&+\dfrac{271360 \zeta (3)}{27}-17600 \zeta (5)+\dfrac{4603055}{81}~,~\vspace{0.2cm}\\
\gamma_4&=n_f^4 \left(-\dfrac{640 \zeta (3)}{243}+\dfrac{64 \pi ^4}{1215}-\dfrac{520}{243}\right)+n_f^3 \left(\dfrac{25696 \zeta (3)}{81}-\dfrac{10240 \zeta (5)}{27}+\dfrac{448 \pi ^4}{405}+\dfrac{91865}{729}\right)\\
&+n_f^2 \left(\dfrac{92800 \zeta (3)^2}{27}+\dfrac{4021648 \zeta (3)}{243}-\dfrac{528080 \zeta (5)}{81}+\dfrac{36800 \pi ^6}{5103}-\dfrac{33260 \pi ^4}{243}+\dfrac{2641484}{729}\right)\\
&+n_f \left(-\dfrac{151360 \zeta (3)^2}{9}-\dfrac{25076032 \zeta (3)}{81}+\dfrac{99752360 \zeta (5)}{243}-\dfrac{3640000 \zeta (7)}{27}+\dfrac{2038742 \pi ^4}{1215}\right.\\
& \left.-\dfrac{150736283}{729}-\dfrac{255200 \pi ^6}{1701}\right)+193600 \zeta (3)^2+\dfrac{92804932 \zeta (3)}{243}+825440 \zeta (7)\\
&-\dfrac{463514320 \zeta (5)}{243}+\dfrac{96800 \pi ^6}{189}+\dfrac{99512327}{81}-\dfrac{698126 \pi ^4}{405}~.
\end{cases}
\end{split}
\end{align}}}
In the Veneziano limit our conventions for the $\beta$ function and its perturbative coefficients are 
\begin{equation}
\beta(\lambda) = \frac{1}{2}\frac{d \lambda}{d\log\mu} = -\sum_{n=0}^\infty \beta^{\rm V}_n \,\lambda^{n+2}~,
\end{equation}
and similarly for the anomalous dimension
\begin{equation}
\gamma(\lambda) =  -\sum_{n=0}^\infty \gamma^{\rm V}_n \,\lambda^{n+2}~.
\end{equation}
These perturbative coefficients in the Veneziano limit can be obtained from \cite{Herzog:2017ohr,Luthe:2017ttg,Chetyrkin:2017bjc} up to five-loop order and they read
{\small{\begin{align}
\begin{split}
\!\!\!\!\!\!\!\!\!\!\!\!\!\!\!\!\!\!\!\!\!\!\!\!\!\!\!\!\!\!\!\!\!\!\!\!\!\begin{cases}
\beta^{\rm V}_0 &=\dfrac{11}{3}-\dfrac{2 x}{3}~,~~~~~~\beta^{\rm V}_1=\dfrac{34}{3}-\dfrac{13 x}{3}~,~~~~~~\beta^{\rm V}_2=\dfrac{56 x^2}{27}-\dfrac{1709 x}{54}+\dfrac{2857}{54}~,~\vspace{0.3cm}\\
\beta^{\rm V}_3&=\dfrac{130 x^3}{243}+x^2 \left(\dfrac{28 \zeta (3)}{3}+\dfrac{8654}{243}\right)-x \left(\dfrac{20 \zeta (3)}{9}+\dfrac{485513}{1944}\right)+\dfrac{44 \zeta (3)}{9}+\dfrac{150473}{486}~,~\vspace{0.3cm}\\
\beta^{\rm V}_4&=x^4 \left(\dfrac{61}{486}-\dfrac{52 \zeta (3)}{81}\right)+x^3 \left(-\dfrac{1937 \zeta (3)}{81}+\dfrac{20 \zeta (5)}{3}+\dfrac{7 \pi ^4}{90}-\dfrac{5173}{432}\right)\\
&+x^2 \left(\dfrac{33125 \zeta (3)}{108}-\dfrac{1630 \zeta (5)}{9}-\dfrac{241 \pi ^4}{540}+\dfrac{3952801}{7776}\right)\\
&+x \left(-\dfrac{231619 \zeta (3)}{648}+\dfrac{4090 \zeta (5)}{9}+\dfrac{77 \pi ^4}{540}-\dfrac{11204369}{5184}\right)\\
&+\dfrac{38851 \zeta (3)}{162}-330 \zeta (5)-\dfrac{121 \pi ^4}{540}+\dfrac{8268479}{3888}\,,
\end{cases}
\end{split}
\end{align}}}
{\small{\begin{align}
\begin{split}
\begin{cases}
\gamma^{\rm V}_0 &=3~,~~~~~~\gamma^{\rm V}_1=\dfrac{203}{12}-\dfrac{5 x}{3}~,~~~~~~\gamma^{\rm V}_2=-\dfrac{35 x^2}{27}-x \left(12 \zeta (3)+\dfrac{1177}{54}\right)+\dfrac{11413}{108}~,~\vspace{0.3cm}\\
\gamma^{\rm V}_3&=x^3 \left(\dfrac{16 \zeta (3)}{9}-\dfrac{83}{81}\right)+x^2 \left(20 \zeta (3)-\dfrac{2 \pi ^4}{15}+\dfrac{899}{162}\right)\\
&+x \left(-\dfrac{889 \zeta (3)}{3}+160 \zeta (5)+\dfrac{11 \pi ^4}{15}-\dfrac{23816}{81}\right)+\dfrac{1157 \zeta (3)}{9}-220 \zeta (5)+\dfrac{460151}{576}~,~\vspace{0.3cm}\\
\gamma^{\rm V}_4&=x^4 \left(-\dfrac{80 \zeta (3)}{81}+\dfrac{8 \pi ^4}{405}-\dfrac{65}{81}\right)+x^3 \left(\dfrac{3278 \zeta (3)}{81}-\dfrac{416 \zeta (5)}{9}+\dfrac{46 \pi ^4}{405}+\dfrac{8029}{486}\right)\\
&+x^2 \left(\dfrac{368 \zeta (3)^2}{3}+\dfrac{35323 \zeta (3)}{54}-\dfrac{778 \zeta (5)}{3}+\dfrac{160 \pi ^6}{567}-\dfrac{2843 \pi ^4}{540}+\dfrac{1315303}{7776}\right)\\
&\!\!\!\!+x \left(-\dfrac{616 \zeta (3)^2}{3}-\dfrac{2598341 \zeta (3)}{648}+5304 \zeta (5)-1820 \zeta (7) +\dfrac{17639 \pi ^4}{810}-\dfrac{1100 \pi ^6}{567}-\dfrac{46120039}{15552}\right)\\
&\!\!\!\!+\dfrac{968 \zeta (3)^2}{3}+\dfrac{215171 \zeta (3)}{162}+3850 \zeta (7)-\dfrac{66235 \zeta (5)}{9}+\dfrac{1210 \pi ^6}{567}-\dfrac{3157 \pi ^4}{405}+\dfrac{29826469}{5184}~.
\end{cases}
\end{split}
\end{align}}}

\section{RG Functions in the Large-$n_f$ Limit}
\label{app:largenf}

In this appendix we review the results for $\beta$ and $\gamma$ at large $n_f$. 
Ref. \cite{Gracey:1996he} showed that in the large-$n_f$ limit of QCD (with $n_c$ kept finite) the $\overline{{\rm MS}}$ $\beta$-function for the 't~Hooft-like coupling $\lambda = n_f a$ can be written 
as\footnote{In the main text the large-$n_f$ coupling is denoted by $\lambda_f$ to distinguish it from the coupling in the Veneziano limit. For simplicity of notation, we will drop the subscript $f$ in this section and denote it simply by $\lambda$.} 
\begin{equation}
\beta(\lambda) = \frac{4 T_F}{3} \lambda^2 +\frac{1}{n_f} \beta^{(1)}(\lambda) +\mathcal{O}\left(n_f^{-2}\right)~,\label{eq:betagrac}
\end{equation}
where $\beta^{(1)}(\lambda)$ satisfies the equation
\begin{equation}
\lambda\frac{d}{d\lambda}\left(\frac{\beta^{(1)}(\lambda)}{\lambda^2}\right)  = f(2 - 4 T_F \lambda/3)~. \label{eq:beta1}\\
\end{equation}
The function $f$ on the right-hand side is
\begin{equation}
f(\nu) \equiv \frac{\Gamma (2 \nu ) \left(C_A \left(\nu ^4-\tfrac92 \nu ^3+11 \nu ^2-\tfrac{45}{4} \nu +\tfrac72\right)-4 \,C_F (\nu -3) (\nu -1) (\nu -\tfrac32) (\nu -\tfrac12)\right)}{3(\nu -1) \Gamma (2-\nu ) \Gamma (\nu )^2 \Gamma (\nu +1)}~.
\end{equation}
In the case of $SU(n_c)$ with fundamental flavors the group theory factors are 
\begin{equation}
C_A = n_c\,,~C_F=\frac{n_c^2-1}{2 n_c}\,,~T_F=\frac 12~,
\end{equation}
while for $U(1)$ theories they are
\be
C_A=0\,,\quad C_F = T_F =1\,.
\ee
From eq. \eqref{eq:beta1} one obtains \cite{Holdom:2010qs}
\begin{equation}
\beta^{(1)}(\lambda) = \lambda^2\left(-\frac{11 C_A}{3} + \int_0^\lambda \frac{d\lambda'}{\lambda'}f(2 - 4 T_F \lambda'/3)\right)~.\label{eq:largenfbeta}
\end{equation}
Note that the small-$\lambda$ expansion of the integral starts from $\mathcal{O}(\lambda)$, and the integration constant $-\frac{11 C_A}{3}$ is fixed to agree with the one-loop $\beta$-function. 
Equation \eqref{eq:largenfbeta} generalizes results obtained previously for QED in \cite{Espriu:1982pb, PalanquesMestre:1983zy, Gracey:1992ns}.
Similarly for the $\overline{{\rm MS}}$ mass anomalous dimension one has \cite{Espriu:1982pb,PalanquesMestre:1983zy}
\begin{equation}
\gamma = \frac{1}{n_f}\gamma^{(1)}(\lambda) +\mathcal{O}\left(n_f^{-2}\right)~,
\label{eq:gammaDefLargenf}
\end{equation}
with
\be
\gamma^{(1)}(\lambda) = g(2-4 T_F\lambda/3)~,\quad\quad
g(\nu)\equiv -\frac{C_F}{2  T_F}\frac{(2 \nu -1) \Gamma (2 \nu )}{\Gamma (\nu )^2 \Gamma (\nu +1) \Gamma (2-\nu )}~.\label{eq:largenfgamma}
\ee
A convenient approach to derive these formulas (though not the one pursued in the first papers on the subject \cite{Espriu:1982pb, PalanquesMestre:1983zy}) is to (formally) continue the theory to dimension $d$ with $2<d<4$, and exploit that in this range of dimensions the large-$n_f$ theory admits a controllable IR fixed point.  The functions $\partial_\lambda \beta=\beta^\prime$ and $\gamma$ at the fixed point at large $n_f$, call them $\beta^{\prime *}$ and $\gamma^*$, are related to observable scaling dimensions of the CFT.  At large-$n_f$ the IR CFT decouples in a non-local mean-field-theory sector associated to the gluons and a free CFT associated to the matter fields. One can then compute these observables in $1/n_f$ conformal perturbation theory.\footnote{Interestingly, the IR fixed point for large-$n_f$ QCD in $2<d<4$ coincides with the UV fixed point for the Thirring model of free fermions deformed by a certain (vector)$^2$-type of quartic interaction \cite{Hasenfratz:1992jv} (see \cite{Hasenfratz:1992jv, Gracey:1996he} for more details about the map between the two theories). Beyond the large-$n_f$ limit, the IR fixed point of QCD becomes weakly coupled in the limit $d\to4$, while the UV fixed point of the Thirring model becomes weakly coupled in the limit $d\to2$. The result \eqref{eq:beta1} for the $\beta$ function was indeed computed in \cite{Gracey:1996he} using the Thirring description of the fixed point, and it agrees with the large-$n_f$ limit of the five-loop QCD $\beta$ function.} We then have that $\beta^{\prime *}=\beta^{\prime *}(d)$ and $\gamma^*=\gamma^*(d)$ are non-trivial functions of the space-time dimension, at each order in the $1/n_f$ expansion.
On the other hand, using the leading order $\beta$-function for large $n_f$ we can compute the critical value of the coupling constant $\lambda^*=\lambda^*(d)$ at the fixed point. Inverting this function and plugging the result
in $\beta^{\prime *}$ and $\gamma^*$ gives us the final answer for $\beta$ and $\gamma$ as functions of $\lambda$.

We now show how to use this approach to obtain the formula \eqref{eq:largenfgamma}. We consider the theory in $d$ dimensions with Euclidean signature. To start with, let us derive the propagator of the gluon in the IR for $n_f = \infty$. At leading order at large $n_f$ the only contribution to the 1PI two-point function is the fermionic bubble, hence the exact propagator in $\xi$ gauge takes the simple form (stripping off the color indices)
\begin{equation}
P_{\mu\nu}(q)= \frac{\lambda}{n_f}\frac{1}{q^2}\frac{1}{1+\lambda  \, T_F C(\nu)\, q^{d-4} } \Pi_{\mu\nu}(q)+ \xi \frac{\lambda}{n_f} \frac{q_\mu q_\nu}{q^4}+\mathcal{O}(n_f^{-2})~,
\end{equation}
where for convenience we have defined $\nu=d/2$, and $\Pi_{\mu\nu}(q)$ is the transverse projector
\begin{equation}
\Pi_{\mu\nu}(q)\equiv \delta_{\mu\nu} - \frac{q_\mu q_\nu}{q^2}~.
\end{equation}
A simple one-loop calculation shows that
\begin{equation}
C(\nu)=\frac{1}{(4\pi)^{\nu}}\frac{4(\nu-1)\Gamma(2-\nu)\Gamma(\nu-1)^2}{(2\nu-1)\Gamma(2\nu-2)}~.
\label{eq:CdDef}
\ee
For simplicity we take in the following the Landau gauge $\xi=0$. The IR limit corresponds to $\lambda \, q^{2\nu-4} \gg 1$, leading to
\begin{equation}
P^{\text{IR}}_{\mu\nu}(q)= \frac{1}{n_f}\frac{1}{T_F C(\nu)\, q^{d-2} } \Pi_{\mu\nu}(q)+\mathcal{O}(n_f^{-2})~.
\end{equation}
Using this propagator, we can compute the correction to the 1PI two-point function of the fermion field, giving
\begin{equation}
\Sigma(p^2) \gamma_\sigma p^\sigma = \frac{1}{n_f}\frac{C_F}{T_F C(\nu)}\int \frac{d^d q}{(2\pi)^d} \frac{(\gamma_\mu\gamma_\rho\gamma_\nu)\Pi^{\mu\nu}(q) (p+q)^\rho}{(p+q)^2 q^{d-2}} +\mathcal{O}(n_f^{-2})~,
\label{eq:SigmaAppB}
\end{equation}
where the indices in the fundamental representation have been contracted on both sides, giving the factor of $C_F$ in the numerator on the r.h.s. of eq.\eqref{eq:SigmaAppB}.
Multiplying both sides by $\gamma_\omega p^\omega$ and taking the trace over the fermionic indices, we obtain
\begin{equation}
\Sigma(p^2) p^2 = \frac{1}{n_f}\frac{C_F}{T_F C(\nu)}\int \frac{d^d q}{(2\pi)^d} \frac{F(p,q)}{(p+q)^2 q^{d-2}} +\mathcal{O}(n_f^{-2})~,
\label{eq:SigmaEx}
\end{equation}
where
\begin{equation}
F(p,q)= \frac{3-d}{2} p^2 -\frac{2-d}{2} q^2 - \frac{(p^2)^2}{2 q^2} + \frac{3-d}{2}(p+q)^2+
\frac{p^2(p+q)^2}{q^2}-\frac{((p+q)^2)^2}{2q^2}~.
\end{equation}
Taking a Wilsonian approach, we will perform the integral in the slice of momenta $b\Lambda < |q| < \Lambda$, where $\Lambda$ is a UV cutoff and $0<b<1$. Expanding around $p=0$, we have
\begin{equation}
\frac{F(p,q)}{(p+q)^2} = \frac{1}{q^2}\left[(1-d) p\cdot q+(3-d)p^2 -2 (2-d) \frac{(p\cdot q)^2}{q^2}\right]+\mathcal{O}(|p|^3 |q|^{-3})~,
\label{eq:FpqExp}
\end{equation}
where we retain orders up to $\mathcal{O}(|p|^2 |q|^{-2})$ because higher orders lead to a UV finite integral. We can now plug eq.(\ref{eq:FpqExp}) inside the integral in eq.(\ref{eq:SigmaEx}) and use $SO(d)$ invariance 
to conclude that the term linear in $q$ cannot contribute, while in the term quadratic in $q$ we can replace
\begin{equation}
q_\mu q_\nu \to\frac{1}{d} q^2 \delta_{\mu\nu}~.
\end{equation}
We then get 
\begin{equation}
\Sigma(p^2=0)= \frac{1}{n_f}\frac{C_F}{T_F C(\nu)} \frac{(d-4)(d-1)\text{Vol}(S^{d-1})}{d (2\pi)^d} \log(b)+\mathcal{O}(n_f^{-2})~.
\end{equation}
This means that in the effective action at the scale $b\Lambda$ we have a non-canonical kinetic term for the fermion, with coefficient 
$Z_\psi = 1 -\Sigma(p^2=0)$.
Similarly, we can compute the correction to the mass due to the modes between $b\Lambda$ and $\Lambda$. The Feynman diagram evaluates to the following integral
\begin{equation}
-(Z_m-1) m \mathds{1}= -m \frac{1}{n_f}\frac{C_F}{T_F C(\nu)}\int \frac{d^d q}{(2\pi)^d} \frac{(\gamma_\mu\gamma_\nu)\Pi^{\mu\nu}(q) }{(p+q)^2 q^{d-2}} +\mathcal{O}(n_f^{-2})~\,,
\end{equation}
whose UV divergence is readily computed as before.
Going to the canonical normalization of the kinetic term, the correction to the renormalization of the mass is
\be
\frac{Z_m}{Z_\psi} 
 =1 -\frac{1}{n_f}\frac{C_F}{T_F C(\nu)} \frac{4 (d-1)\text{Vol}(S^{d-1})}{d (2\pi)^d}\log(b) +\mathcal{O}(n_f^{-2})~.
\ee
Using eq.(\ref{eq:CdDef}) and $\text{Vol}(S^{d-1}) = 2\pi^{\nu}/\Gamma(\nu)$ we then have
\begin{equation}
\gamma(\nu) = \frac{d(\log(Z_m/Z_\psi))}{d\log(b)} = \frac{1}{n_f} g(\nu) +\mathcal{O}(n_f^{-2})\,, 
\label{eq:gammanu}
\ee
where $g(\nu)$ is exactly the function defined in eq. \eqref{eq:largenfgamma}.

In order to find the dependence of the fixed point coupling $\lambda^*$ on $\nu$, we use the $\epsilon$-expansion around $d=4$. 
The beta function 
in $d=4-2\epsilon$ is one-loop exact at leading order in $1/n_f$:
\begin{equation}
\beta(\lambda)= - \epsilon \lambda + \frac{4 T_F}{3}\lambda^2 +\mathcal{O}(n_f^{-1})~.
\end{equation}
For $\epsilon>0$ we get a real IR fixed point at 
\begin{equation}
\lambda^* = \frac{3}{4 T_F}\epsilon + \mathcal{O}(n_f^{-1})\rightarrow \nu(\lambda^*) = 2-\frac {4 T_F}{3} \lambda^*+ \mathcal{O}(n_f^{-1}).
\end{equation}
Plugging $\nu(\lambda^*)$ in eq.(\ref{eq:gammanu}) and replacing $\lambda^*\rightarrow \lambda$ finally reproduces eq.(\ref{eq:gammaDefLargenf}).
The same strategy works also for the $\beta$ function, in which case one needs to consider the scaling dimension of the gauge kinetic operator.
We refer the reader to \cite{Gracey:1996he}  for this case, which is more complicated since it involves 2-loop diagrams. 

\section{Details on the Numerical Resummation}
\label{app:numerics}

We approximate the Borel transform of the RG functions $\beta$ and $\gamma$ using Pad\'e - approximants.\footnote{More refined analysis can be made when some analytic properties of the Borel function are known or assumed
and more perturbative coefficients are known. See e.g. section 4 of \cite{Serone:2018gjo} for an application in the 2d $\lambda \phi^4$ theory.}
Given the asymptotic series for a function $f(a)$ 
\be
a^{\zeta_f} \sum_{n=0}^{\infty} f_{n} a^{n}\,,
\label{Betagn}
\ee
we define the generalized Borel-Le~Roy transform as
\be
{\cal B}_f(t) = \sum_{n=0}^\infty \frac{f_n}{\Gamma(n+b+1)} t^n\,,\quad \quad 
\label{LeRoy}
\ee
where $b>-1$ is a real parameter.\footnote{ The ordinary Borel function corresponds to $b=0$.} 
The parameter $\zeta_f$ in eq.(\ref{Betagn}) is chosen in such a way that $f_0\neq 0$, so we have $\zeta_f=2$ for $\beta$ and $\zeta_f=1$ for both $\gamma$ and $\gamma_g\propto \beta^\prime$.
The Borel resummed function is obtained as
\be
f_B(a) = a^{\zeta_f} \int_0^\infty dt \, t^b e^{-t} {\cal B}_f(a t)\,.
\label{Betag}
\ee
Given the first $N+1$ terms of the series expansion of the Borel function (\ref{LeRoy}), its $[m/n]$ Pad\'e approximation reads
\be
{\cal B}_b^{[m/n]}(t) = \frac{\sum_{p=0}^m c_p(b) t^p}{1+\sum_{q=1}^n d_q(b) t^q}\,,
\label{pade2}
\ee
with $m+n = N$. The $m+n+1$ $b$-dependent coefficients $c_p$ and $d_q$ are determined by expanding eq.~(\ref{pade2}) around $t=0$  and matching the result up to the $t^{N}$ term in eq.~(\ref{LeRoy}).
Plugging eq.~(\ref{pade2}) in eq.~(\ref{Betag}) leads to an approximation of the function $f_B(a)$ given by
\be
f_B^{[m/n]}(a) = a^{\zeta_f} \int_0^\infty \!\!dt \, t^b e^{-t} {\cal B}_b^{[m/n]}(a t)\,.
\label{FgPB}
\ee
For the values of $a$ where $f_B(a)$ provides a good approximation of the exact function $f(a)$, 
$f_B(a)$ should be independent of the dummy variable $b$, but a dependence will remain in its Pad\'e approximation $f_B^{[m/n]}(a)$.
Such a dependence can be used to estimate how well we are approximating the exact Borel function.
We have used the large-$n_f$ limit to find which range of values for $b$ gives more accurate results. It turns out that largish values of $b$
are preferred, and we have taken $b_0=10$ as our central value.

At sufficiently high order, the $m+n$ zeroes and poles of the Pad\'e $[m/n]$ approximants can be used to ``reconstruct" the analytic property of the Borel function, since most of them accumulate
on specific branch-cuts of the function, see appendix \ref{app:pade} for a more precise statement and e.g. fig.2 of \cite{Marino:2019eym} or fig.4 of \cite{Serone:2017nmd} for examples in 
2d or 1d theories. In principle one could see the possible appearance of IR renormalon singularities in this way. However the occurrence of random spurious poles hinder any possible use of the location of zeros and poles of the approximants  when few terms are available, like in our case. When a pole occurs in the positive real axis, independently of its possible interpretation, we  take the Cauchy principal value of the
integral and include the ambiguity, given by the residue at that pole, into the final error estimate.
 
It is notoriously hard to find a systematic error estimate in numerical resummations, which are always subject to some arbitrariness.
We define our (non rigorous) error $\Delta^{[m/n]}$ as a sum of three contributions
\be
 \Delta \, f_{B}^{[m/n]} \equiv \Delta^{[m/n]}= \Delta_b^{[m/n]}  + \Delta_r^{[m/n]}  + \Delta_{{\rm np}}   \,.
 \label{SumErrors}
\ee 
The factor $\Delta_b$ measures the error due to the choice of the $b$ parameter and is taken as follows:
\be
 \Delta_b^{[m/n]}  =  \frac 12\Big| \underset{b\in \mathcal{B}}{\rm max}f_B^{[m/n]}(b) -  \underset{b\in \mathcal{B}}{\rm min}f_B^{[m/n]}(b) \Big| \,. \quad 
\label{errorPB1}
\ee 
$\mathcal{B}$ is a grid of values of the parameter $b$ inside the interval $[b_0-\Delta b,b_0+\Delta b]$, where the value of $\Delta b$ is an arbitrary choice. We have conservatively taken $\Delta b = 10$, and the spacing of the grid to be $2$, so that $\mathcal{B}=\{2k|k\in\mathbb{N},\, 0\leq k\leq 10\}$. 
The factor $\Delta_r$ is non-vanishing only if the approximant has poles in the positive real axis, in which case we have
\be
 \Delta_r^{[m/n]}  = a^{\zeta_f-1}  \sum_p {\rm Res}\bigg|  \Big(\frac{u_p}{a}\Big)^b e^{-u_p/a}  {\cal B}_{b_0}^{[m/n]}(u_p)\ \bigg| \,, \quad 
\label{errorPB2}
\ee 
where ${\rm Res}$ indicates the residue of the function, $u_p$ is the location of the pole and $p$ runs over the number of poles in the positive real $u=t a$ plane.
The factor $\Delta_{{\rm np}}$ represents the theoretical error due to possible non-perturbative contribution to the RG functions. 
As an order-of-magnitude estimate it is taken as
\be
 \Delta_{{\rm np}}  = c_{{\rm np}} \, e^{-\frac{\kappa}{a}} \,, \quad 
\label{errorNP}
\ee 
where $c_{{\rm np}}$ is an arbitrary coefficient, $\kappa={\rm Min} (1,1/ \beta_0)$ in QCD, $\kappa=1 /\beta_0$ in the Veneziano limit, 1 being the position of the instanton anti-instanton singularity in the Borel plane, and $1 /\beta_0$ being the exponent of the leading non-perturbative correction in eq. \eqref{eq:transseries}, as explained in section \ref{sec:gce}.
The error $\Delta_{{\rm np}}$ applies for the series of $\beta$, $\gamma$ and $\gamma_g$.

A similar analysis applies in the conformal expansion where the coupling $a$ is replaced by $\epsilon$ defined in eq. \eqref{eq:epsDef}.
As can easily be seen from eqs.\eqref{eq:astar} and \eqref{eq:gammastar}, in the conformal case $\zeta_f=1$ for both $a^*$ and $\gamma^*$ while $\zeta_f=2$ for $\gamma_g^*$, and all considerations made apply. In particular, the theoretical error is taken as in eq.(\ref{errorNP}), where $\beta_0$ is proportional to $\epsilon$ and $a$ is replaced by $a^*(\epsilon)$.

\section{Convergence of Pad\'e Approximants}
\label{app:pade}

This appendix briefly discusses  what is known about the convergence properties of $[m/n]$ Pad\'e approximants in general and can be read independently of the main text.

Given an analytic function $f(z)$, we would like to understand if and in what sense its approximants $f^{[m/n]}(z)$ converge to $f(z)$ when $m,n\rightarrow \infty$.
It is clear intuitively that it is difficult to prove point-wise convergence for Pad\'e-approximants, even for single-valued meromorphic functions. 
Indeed, suppose that $f(z)$ is meromorphic in some domain ${\cal D}$ of the complex plane that includes the origin and has $n$ poles of order $\alpha_i\in {\cal D}$, 
so that the total pole multiplicity is $M=\sum_{i=1}^n \alpha_i$. The approximant $f^{[m/n]}(z)$ has instead $n$ poles in the complex plane and for sufficiently large $n$ the total multiplicity
of the poles of $f^{[m/n]}(z)$ in ${\cal D}$ can exceed $M$. There are then at least a finite set of points where $f^{[m/n]}(z)$ diverge while $f(z)$ is finite. 
The most general results on the convergence of Pad\'e approximants are based on convergence in ``capacity". 
There are various equivalent ways to define it, see e.g. chapter 6.6 of \cite{baker1996pade}. The name arises from the definition in potential theory, where it admits an interpretation in terms of electrostatic capacity.
Capacity is a property of a set that essentially measures its magnitude. The capacity of a finite two-dimensional domain is non-vanishing. 
Codimension one regions such as a circle or a segment, despite being of measure zero in $\mathbb{C}$, have also non-zero capacity. For example, a disk of radius $R$ and a circle of radius R (the boundary of the disk) have the same capacities: ${\rm cap}(D_R) = {\rm cap}(S^1_R)= R$. The capacity of any countable number of points is zero. 

The most general results on the convergence of Pad\'e approximants are due to Stahl \cite{STAHL1997139}, building on previous works by Nuttall, and are  based on convergence in capacity.
Following Stahl, we denote by $[m/n](z)\equiv f^{[m/n]}(1/z)$ the Pad\'e approximants of $f(z)$ expanded around $z=\infty$. The function $f(z)$ is assumed to be analytic
at infinity, all its singularities are in a compact set $E\subseteq \overline{\mathbb{C}}$  with ${\rm cap}(E)=0$, and it has analytic continuations (not necessarily single-valued) along any path on $\overline{\mathbb{C}\!}\setminus\! E$, where 
$\overline{\mathbb{C}}$ is the complex plane $\mathbb{C}$ extended with the point at infinity, i.e. the Riemann sphere. 
Under these assumptions, Stahl has shown that there exists a unique maximal domain of convergence ${\cal D}_f$ such that, for every compact set $V\subseteq {\cal D}_f$ and for every $\epsilon>0$,
\be
\lim_{m,n\rightarrow \infty} {\rm cap} \Big(z\in V| \,  |f(z) - [m/n](z)| > \epsilon  \Big) = 0 \,.
\label{eq:Stahl1}
\ee
The Pad\'e approximants $[m/n]$ should be parametrically diagonal, namely
\be
\lim_{m,n\rightarrow \infty}\frac{m}{n} = 1\,.
\ee
The domain ${\cal D}_f$ is determined as the one whose boundary $\partial {\cal D}_f$ has minimum capacity among all the domains including the point at infinity and where $f(z)$ has a single-valued analytic continuation. For a meromorphic function ${\cal D}_f =  \overline{\mathbb{C}}$, while ${\cal D}_f \subset  \overline{\mathbb{C}}$ if $f$ has branch-point singularities. The boundaries  $\partial {\cal D}_f$ of the domain should be thought as the various branch-cuts connecting branch-point singularities and ${\cal D}_f $ as the unique domain delimited by branch-cuts with minimum capacity where $f(z)$ is single-valued.
The complement set $F=\overline{\mathbb{C}}\setminus {\cal D}_f$ has empty interior and corresponds to a compact set of zero capacity plus, if $f(z)$ has branch-points,
an union of arcs in $\mathbb{C}$. For single-valued functions the convergence in capacity over $\overline{\mathbb{C}}\!\setminus\! E$ can be shown to occur for arbitrary approximants, not necessarily parametrically diagonal. We refer the reader to \cite{STAHL1997139} for more details and other results based on different assumptions and to \cite{baker1996pade} for a more comprehensive introduction to the subject.

Interestingly enough, Stahl has also proved that almost all the poles and zeros of the approximant $[m/n](z)$ cluster to the set $F$. Isolated spurious poles in ${\cal D}_f$ can however occur and generally hinder
a point-wise convergence. For functions with a single branch-cut, the set $F$ (modulo a capacity zero set) coincides with the branch-cut. It is straightforward to check how zeros and poles of  large order approximants of simple known functions cluster on their branch-cut, as expected. For  more complicated functions  with multiple branch-cuts the clustering of zeros and poles allows us to ``reconstruct" the set $F$.

It is important to emphasize that essentially all results on Pad\'e approximants assume that the original function to be approximated is analytic at the point where it is expanded.\footnote{More precisely, 
the point could also be singular, but the singularity should not be visible in some Riemann sheet, where locally the function admits a convergent power expansion. 
A simple example of this kind is $f(z) = \log(1-z)/z$. This function is analytic at $z=0$ only in one sheet but develops a simple pole in other sheets.} 
This is not the case for functions $f$ that have a divergent asymptotic series, so the convergence of Pad\'e approximants of $f$ cannot be established.
On the other hand, by construction the Borel transform ${\cal B}f$ of $f$ is analytic around the point of expansion and the convergence theorems might apply.
Moreover, since the original function is recovered by an integral, convergence in capacity (which is stronger than convergence in measure) is enough to establish 
the point-wise convergence in $f$, as long as the real positive axis is included in ${\cal D}_{{\cal B}f}$ and the Laplace transform is well defined.

\bibliographystyle{JHEP}
\bibliography{CW_QCD_v5}

\providecommand{\href}[2]{#2}\begingroup\raggedright\begin{thebibliography}{10}

\bibitem{Appelquist:1996dq}
T.~Appelquist, J.~Terning, and L.~C.~R. Wijewardhana, {\it {The Zero
  temperature chiral phase transition in SU(N) gauge theories}},  {\em Phys.
  Rev. Lett.} {\bf 77} (1996) 1214--1217,
  [\href{http://arxiv.org/abs/hep-ph/9602385}{{\tt hep-ph/9602385}}].

\bibitem{Yamawaki:1985zg}
K.~Yamawaki, M.~Bando, and K.-i. Matumoto, {\it {Scale Invariant Technicolor
  Model and a Technidilaton}},  {\em Phys. Rev. Lett.} {\bf 56} (1986) 1335.

\bibitem{Appelquist:1988yc}
T.~Appelquist, K.~D. Lane, and U.~Mahanta, {\it {On the Ladder Approximation
  for Spontaneous Chiral Symmetry Breaking}},  {\em Phys. Rev. Lett.} {\bf 61}
  (1988) 1553.

\bibitem{Cohen:1988sq}
A.~G. Cohen and H.~Georgi, {\it {Walking Beyond the Rainbow}},  {\em Nucl.
  Phys.} {\bf B314} (1989) 7--24.

\bibitem{Gies:2005as}
H.~Gies and J.~Jaeckel, {\it {Chiral phase structure of QCD with many
  flavors}},  {\em Eur. Phys. J.} {\bf C46} (2006) 433--438,
  [\href{http://arxiv.org/abs/hep-ph/0507171}{{\tt hep-ph/0507171}}].

\bibitem{Gukov:2016tnp}
S.~Gukov, {\it {RG Flows and Bifurcations}},  {\em Nucl. Phys.} {\bf B919}
  (2017) 583--638, [\href{http://arxiv.org/abs/1608.06638}{{\tt
  arXiv:1608.06638}}].

\bibitem{Kuipers:2018lux}
F.~Kuipers, U.~Gursoy, and Y.~Kuznetsov, {\it {Bifurcations in the RG-flow of
  QCD}},  {\em JHEP} {\bf 07} (2019) 075,
  [\href{http://arxiv.org/abs/1812.05179}{{\tt arXiv:1812.05179}}].

\bibitem{Kusafuka:2011fd}
Y.~Kusafuka and H.~Terao, {\it {Fixed point merger in the SU(N) gauge beta
  functions}},  {\em Phys. Rev.} {\bf D84} (2011) 125006,
  [\href{http://arxiv.org/abs/1104.3606}{{\tt arXiv:1104.3606}}].

\bibitem{Jarvinen:2011qe}
M.~Jarvinen and E.~Kiritsis, {\it {Holographic Models for QCD in the Veneziano
  Limit}},  {\em JHEP} {\bf 03} (2012) 002,
  [\href{http://arxiv.org/abs/1112.1261}{{\tt arXiv:1112.1261}}].

\bibitem{Alvares:2012kr}
R.~Alvares, N.~Evans, and K.-Y. Kim, {\it {Holography of the Conformal
  Window}},  {\em Phys. Rev.} {\bf D86} (2012) 026008,
  [\href{http://arxiv.org/abs/1204.2474}{{\tt arXiv:1204.2474}}].

\bibitem{Ryttov:2016ner}
T.~A. Ryttov and R.~Shrock, {\it {Infrared Zero of $\beta$ and Value of
  $\gamma_m$ for an SU(3) Gauge Theory at the Five-Loop Level}},  {\em Phys.
  Rev.} {\bf D94} (2016), no.~10 105015,
  [\href{http://arxiv.org/abs/1607.06866}{{\tt arXiv:1607.06866}}].

\bibitem{Antipin:2018asc}
O.~Antipin, A.~Maiezza, and J.~C. Vasquez, {\it {Resummation in QFT with Meijer
  G-functions}},  {\em Nucl. Phys.} {\bf B941} (2019) 72--90,
  [\href{http://arxiv.org/abs/1807.05060}{{\tt arXiv:1807.05060}}].

\bibitem{Kim:2020yvr}
B.~S. Kim, D.~K. Hong, and J.-W. Lee, {\it {Into the conformal window:
  multi-representation gauge theories}},
  \href{http://arxiv.org/abs/2001.02690}{{\tt arXiv:2001.02690}}.

\bibitem{Ryttov:2017lkz}
T.~A. Ryttov and R.~Shrock, {\it {Physics of the non-Abelian Coulomb phase:
  Insights from Pad{\'e} approximants}},  {\em Phys. Rev.} {\bf D97} (2018),
  no.~2 025004, [\href{http://arxiv.org/abs/1710.06944}{{\tt
  arXiv:1710.06944}}].

\bibitem{Simmons-Duffin:2016gjk}
D.~Simmons-Duffin, {\it {The Conformal Bootstrap}},  in {\em {Proceedings,
  Theoretical Advanced Study Institute in Elementary Particle Physics: New
  Frontiers in Fields and Strings (TASI 2015): Boulder, CO, USA, June 1-26,
  2015}}, pp.~1--74, 2017.
\newblock \href{http://arxiv.org/abs/1602.07982}{{\tt arXiv:1602.07982}}.

\bibitem{Poland:2018epd}
D.~Poland, S.~Rychkov, and A.~Vichi, {\it {The Conformal Bootstrap: Theory,
  Numerical Techniques, and Applications}},  {\em Rev. Mod. Phys.} {\bf 91}
  (2019) 015002, [\href{http://arxiv.org/abs/1805.04405}{{\tt
  arXiv:1805.04405}}].

\bibitem{DeGrand:2015zxa}
T.~DeGrand, {\it {Lattice tests of beyond Standard Model dynamics}},  {\em Rev.
  Mod. Phys.} {\bf 88} (2016) 015001,
  [\href{http://arxiv.org/abs/1510.05018}{{\tt arXiv:1510.05018}}].

\bibitem{Hasenfratz:2018wpq}
A.~Hasenfratz, C.~Rebbi, and O.~Witzel, {\it {Determination of the N$_f$=12
  step scaling function using Mobius domain wall fermions}},  {\em PoS} {\bf
  LATTICE2018} (2019) 306, [\href{http://arxiv.org/abs/1810.05176}{{\tt
  arXiv:1810.05176}}].

\bibitem{Fodor:2018uih}
Z.~Fodor, K.~Holland, J.~Kuti, D.~Nogradi, and C.~H. Wong, {\it {Is SU(3) gauge
  theory with 13 massless flavors conformal?}},  {\em PoS} {\bf LATTICE2018}
  (2018) 198, [\href{http://arxiv.org/abs/1811.05024}{{\tt arXiv:1811.05024}}].

\bibitem{Hasenfratz:2019dpr}
A.~Hasenfratz, C.~Rebbi, and O.~Witzel, {\it {Gradient flow step-scaling
  function for SU(3) with twelve flavors}},  {\em Phys. Rev.} {\bf D100}
  (2019), no.~11 114508, [\href{http://arxiv.org/abs/1909.05842}{{\tt
  arXiv:1909.05842}}].

\bibitem{Baikov:2016tgj}
P.~A. Baikov, K.~G. Chetyrkin, and J.~H. Kuhn, {\it {Five-Loop Running of the
  QCD coupling constant}},  {\em Phys. Rev. Lett.} {\bf 118} (2017), no.~8
  082002, [\href{http://arxiv.org/abs/1606.08659}{{\tt arXiv:1606.08659}}].

\bibitem{Herzog:2017ohr}
F.~Herzog, B.~Ruijl, T.~Ueda, J.~A.~M. Vermaseren, and A.~Vogt, {\it {The
  five-loop beta function of Yang-Mills theory with fermions}},  {\em JHEP}
  {\bf 02} (2017) 090, [\href{http://arxiv.org/abs/1701.01404}{{\tt
  arXiv:1701.01404}}].

\bibitem{Luthe:2017ttg}
T.~Luthe, A.~Maier, P.~Marquard, and Y.~Schroder, {\it {The five-loop Beta
  function for a general gauge group and anomalous dimensions beyond Feynman
  gauge}},  {\em JHEP} {\bf 10} (2017) 166,
  [\href{http://arxiv.org/abs/1709.07718}{{\tt arXiv:1709.07718}}].

\bibitem{Chetyrkin:2017bjc}
K.~G. Chetyrkin, G.~Falcioni, F.~Herzog, and J.~A.~M. Vermaseren, {\it
  {Five-loop renormalisation of QCD in covariant gauges}},  {\em JHEP} {\bf 10}
  (2017) 179, [\href{http://arxiv.org/abs/1709.08541}{{\tt arXiv:1709.08541}}].
  [Addendum: JHEP12,006(2017)].

\bibitem{Banks:1981nn}
T.~Banks and A.~Zaks, {\it {On the Phase Structure of Vector-Like Gauge
  Theories with Massless Fermions}},  {\em Nucl. Phys.} {\bf B196} (1982)
  189--204.

\bibitem{Kaplan:2009kr}
D.~B. Kaplan, J.-W. Lee, D.~T. Son, and M.~A. Stephanov, {\it {Conformality
  Lost}},  {\em Phys. Rev.} {\bf D80} (2009) 125005,
  [\href{http://arxiv.org/abs/0905.4752}{{\tt arXiv:0905.4752}}].

\bibitem{Broadhurst:1992si}
D.~J. Broadhurst, {\it {Large N expansion of QED: Asymptotic photon propagator
  and contributions to the muon anomaly, for any number of loops}},  {\em Z.
  Phys.} {\bf C58} (1993) 339--346.

\bibitem{Baikov:2014qja}
P.~A. Baikov, K.~G. Chetyrkin, and J.~H. Kuhn, {\it {Quark Mass and Field
  Anomalous Dimensions to ${\cal O}(\alpha_s^5)$}},  {\em JHEP} {\bf 10} (2014)
  076, [\href{http://arxiv.org/abs/1402.6611}{{\tt arXiv:1402.6611}}].

\bibitem{Luthe:2016xec}
T.~Luthe, A.~Maier, P.~Marquard, and Y.~Schroder, {\it {Five-loop quark mass
  and field anomalous dimensions for a general gauge group}},  {\em JHEP} {\bf
  01} (2017) 081, [\href{http://arxiv.org/abs/1612.05512}{{\tt
  arXiv:1612.05512}}].

\bibitem{Baikov:2017ujl}
P.~A. Baikov, K.~G. Chetyrkin, and J.~H. Kuhn, {\it {Five-loop fermion
  anomalous dimension for a general gauge group from four-loop massless
  propagators}},  {\em JHEP} {\bf 04} (2017) 119,
  [\href{http://arxiv.org/abs/1702.01458}{{\tt arXiv:1702.01458}}].

\bibitem{Beneke:1998ui}
M.~Beneke, {\it {Renormalons}},  {\em Phys. Rept.} {\bf 317} (1999) 1--142,
  [\href{http://arxiv.org/abs/hep-ph/9807443}{{\tt hep-ph/9807443}}].

\bibitem{Marino:2019fvu}
M.~{Mari\~no} and T.~Reis, {\it {A new renormalon in two dimensions}},
  \href{http://arxiv.org/abs/1912.06228}{{\tt arXiv:1912.06228}}.

\bibitem{Benini:2019dfy}
F.~Benini, C.~Iossa, and M.~Serone, {\it {Conformality Loss, Walking, and 4D
  Complex Conformal Field Theories at Weak Coupling}},  {\em Phys. Rev. Lett.}
  {\bf 124} (2020), no.~5 051602, [\href{http://arxiv.org/abs/1908.04325}{{\tt
  arXiv:1908.04325}}].

\bibitem{Gubser:2002vv}
S.~S. Gubser and I.~R. Klebanov, {\it {A Universal result on central charges in
  the presence of double trace deformations}},  {\em Nucl. Phys.} {\bf B656}
  (2003) 23--36, [\href{http://arxiv.org/abs/hep-th/0212138}{{\tt
  hep-th/0212138}}].

\bibitem{Gorbenko:2018ncu}
V.~Gorbenko, S.~Rychkov, and B.~Zan, {\it {Walking, Weak first-order
  transitions, and Complex CFTs}},  {\em JHEP} {\bf 10} (2018) 108,
  [\href{http://arxiv.org/abs/1807.11512}{{\tt arXiv:1807.11512}}].

\bibitem{Caswell:1974gg}
W.~E. Caswell, {\it {Asymptotic Behavior of Nonabelian Gauge Theories to Two
  Loop Order}},  {\em Phys. Rev. Lett.} {\bf 33} (1974) 244.

\bibitem{Brodsky:2000cr}
S.~J. Brodsky, E.~Gardi, G.~Grunberg, and J.~Rathsman, {\it {Disentangling
  running coupling and conformal effects in QCD}},  {\em Phys. Rev.} {\bf D63}
  (2001) 094017, [\href{http://arxiv.org/abs/hep-ph/0002065}{{\tt
  hep-ph/0002065}}].

\bibitem{Gardi:2001wg}
E.~Gardi and G.~Grunberg, {\it {Conformal expansions and renormalons}},  {\em
  Phys. Lett.} {\bf B517} (2001) 215--221,
  [\href{http://arxiv.org/abs/hep-ph/0107300}{{\tt hep-ph/0107300}}].

\bibitem{Brezin:1976vw}
E.~Brezin, J.~C. Le~Guillou, and J.~Zinn-Justin, {\it {Perturbation Theory at
  Large Order. 1. The $\varphi^{2N}$ Interaction}},  {\em Phys. Rev.} {\bf D15}
  (1977) 1544--1557.

\bibitem{Wilson:1971dc}
K.~G. Wilson and M.~E. Fisher, {\it {Critical exponents in 3.99 dimensions}},
  {\em Phys. Rev. Lett.} {\bf 28} (1972) 240--243.

\bibitem{Kompaniets:2017yct}
M.~V. Kompaniets and E.~Panzer, {\it {Minimally subtracted six loop
  renormalization of $O(n)$-symmetric $\phi^4$ theory and critical exponents}},
   {\em Phys. Rev.} {\bf D96} (2017), no.~3 036016,
  [\href{http://arxiv.org/abs/1705.06483}{{\tt arXiv:1705.06483}}].

\bibitem{Sokal:1980ey}
A.~D. Sokal, {\it {An improvement of Watson's theorem on Borel summability}},
  {\em J. Math. Phys.} {\bf 21} (1980) 261--263.

\bibitem{Aniceto:2018bis}
I.~Aniceto, G.~Basar, and R.~Schiappa, {\it {A Primer on Resurgent Transseries
  and Their Asymptotics}},  {\em Phys. Rept.} {\bf 809} (2019) 1--135,
  [\href{http://arxiv.org/abs/1802.10441}{{\tt arXiv:1802.10441}}].

\bibitem{Carosso:2018bmz}
A.~Carosso, A.~Hasenfratz, and E.~T. Neil, {\it {Nonperturbative
  Renormalization of Operators in Near-Conformal Systems Using Gradient
  Flows}},  {\em Phys. Rev. Lett.} {\bf 121} (2018), no.~20 201601,
  [\href{http://arxiv.org/abs/1806.01385}{{\tt arXiv:1806.01385}}].

\bibitem{Aoki:2016yrm}
Y.~Aoki et~al., {\it {Topological observables in many-flavour QCD}},  {\em PoS}
  {\bf LATTICE2015} (2016) 214, [\href{http://arxiv.org/abs/1601.04687}{{\tt
  arXiv:1601.04687}}].

\bibitem{Cheng:2013eu}
A.~Cheng, A.~Hasenfratz, G.~Petropoulos, and D.~Schaich, {\it {Scale-dependent
  mass anomalous dimension from Dirac eigenmodes}},  {\em JHEP} {\bf 07} (2013)
  061, [\href{http://arxiv.org/abs/1301.1355}{{\tt arXiv:1301.1355}}].

\bibitem{Lombardo:2014pda}
M.~P. Lombardo, K.~Miura, T.~J. Nunes~da Silva, and E.~Pallante, {\it {On the
  particle spectrum and the conformal window}},  {\em JHEP} {\bf 12} (2014)
  183, [\href{http://arxiv.org/abs/1410.0298}{{\tt arXiv:1410.0298}}].

\bibitem{Cheng:2013xha}
A.~Cheng, A.~Hasenfratz, Y.~Liu, G.~Petropoulos, and D.~Schaich, {\it {Finite
  size scaling of conformal theories in the presence of a near-marginal
  operator}},  {\em Phys. Rev.} {\bf D90} (2014), no.~1 014509,
  [\href{http://arxiv.org/abs/1401.0195}{{\tt arXiv:1401.0195}}].

\bibitem{Aoki:2012eq}
Y.~Aoki, T.~Aoyama, M.~Kurachi, T.~Maskawa, K.-i. Nagai, H.~Ohki, A.~Shibata,
  K.~Yamawaki, and T.~Yamazaki, {\it {Lattice study of conformality in
  twelve-flavor QCD}},  {\em Phys. Rev.} {\bf D86} (2012) 054506,
  [\href{http://arxiv.org/abs/1207.3060}{{\tt arXiv:1207.3060}}].

\bibitem{Appelquist:2011dp}
T.~Appelquist, G.~T. Fleming, M.~F. Lin, E.~T. Neil, and D.~A. Schaich, {\it
  {Lattice Simulations and Infrared Conformality}},  {\em Phys. Rev.} {\bf D84}
  (2011) 054501, [\href{http://arxiv.org/abs/1106.2148}{{\tt
  arXiv:1106.2148}}].

\bibitem{Hasenfratz:2016dou}
A.~Hasenfratz and D.~Schaich, {\it {Nonperturbative $\beta$ function of
  twelve-flavor SU(3) gauge theory}},  {\em JHEP} {\bf 02} (2018) 132,
  [\href{http://arxiv.org/abs/1610.10004}{{\tt arXiv:1610.10004}}].

\bibitem{Ryttov:2017kmx}
T.~A. Ryttov and R.~Shrock, {\it {Higher-order scheme-independent series
  expansions of $\gamma_{\bar\psi\psi,IR}$ and $\beta'_{IR}$ in conformal field
  theories}},  {\em Phys. Rev.} {\bf D95} (2017), no.~10 105004,
  [\href{http://arxiv.org/abs/1703.08558}{{\tt arXiv:1703.08558}}].

\bibitem{Gracey:1996he}
J.~A. Gracey, {\it {The QCD Beta function at O($1/N_f$)}},  {\em Phys. Lett.}
  {\bf B373} (1996) 178--184, [\href{http://arxiv.org/abs/hep-ph/9602214}{{\tt
  hep-ph/9602214}}].

\bibitem{Holdom:2010qs}
B.~Holdom, {\it {Large N flavor beta-functions: a recap}},  {\em Phys. Lett.}
  {\bf B694} (2011) 74--79, [\href{http://arxiv.org/abs/1006.2119}{{\tt
  arXiv:1006.2119}}].

\bibitem{Espriu:1982pb}
D.~Espriu, A.~Palanques-Mestre, P.~Pascual, and R.~Tarrach, {\it {The $\gamma$
  Function in the 1/$N_f$ Expansion}},  {\em Z. Phys.} {\bf C13} (1982) 153.

\bibitem{PalanquesMestre:1983zy}
A.~Palanques-Mestre and P.~Pascual, {\it {The 1/$N_f$ Expansion of the $\gamma$
  and Beta Functions in {QED}}},  {\em Commun. Math. Phys.} {\bf 95} (1984)
  277.

\bibitem{Gracey:1992ns}
J.~A. Gracey, {\it {Algorithm for computing the beta function of quantum
  electrodynamics in the large N(f) expansion}},  {\em Int. J. Mod. Phys.} {\bf
  A8} (1993) 2465--2486, [\href{http://arxiv.org/abs/hep-th/9301123}{{\tt
  hep-th/9301123}}].

\bibitem{Hasenfratz:1992jv}
A.~Hasenfratz and P.~Hasenfratz, {\it {The Equivalence of the SU(N) Yang-Mills
  theory with a purely fermionic model}},  {\em Phys. Lett.} {\bf B297} (1992)
  166--170, [\href{http://arxiv.org/abs/hep-lat/9207017}{{\tt
  hep-lat/9207017}}].

\bibitem{Serone:2018gjo}
M.~Serone, G.~Spada, and G.~Villadoro, {\it {$\lambda \phi^4$ Theory I: The
  Symmetric Phase Beyond NNNNNNNNLO}},  {\em JHEP} {\bf 08} (2018) 148,
  [\href{http://arxiv.org/abs/1805.05882}{{\tt arXiv:1805.05882}}].

\bibitem{Marino:2019eym}
M.~Mari{\~n}o and T.~Reis, {\it {Renormalons in integrable field theories}},
  \href{http://arxiv.org/abs/1909.12134}{{\tt arXiv:1909.12134}}.

\bibitem{Serone:2017nmd}
M.~Serone, G.~Spada, and G.~Villadoro, {\it {The Power of Perturbation
  Theory}},  {\em JHEP} {\bf 05} (2017) 056,
  [\href{http://arxiv.org/abs/1702.04148}{{\tt arXiv:1702.04148}}].

\bibitem{baker1996pade}
G.~A. Baker~Jr. and P.~Peter Graves-Morris, {\em Pad{\'e} Approximants}.
\newblock Encyclopedia of Mathematics and its Applications. Cambridge
  University Press, 1996.

\bibitem{STAHL1997139}
H.~Stahl, {\it {The convergence of Pad{\'e} approximants to functions with
  branch points}},  {\em Journal of Approximation Theory} {\bf 91} (1997),
  no.~2 139 -- 204.

\end{thebibliography}\endgroup

\end{document}